\documentclass[twocolumn]{aastex62}

\usepackage{rotating}

\usepackage{natbib}
\usepackage{amsmath,amssymb}
\usepackage{graphicx}
\usepackage{dcolumn}
\usepackage{bm}

\newcommand{\beq}{\begin{equation}}
\newcommand{\eeq}{\end{equation}}

\begin{document}

\title{Extended Cyclotron Resonant Heating of the Turbulent Solar Wind}


\author{Trevor A. Bowen}\email{tbowen@berkeley.edu}
\affiliation{Space Sciences Laboratory, University of California, Berkeley, CA 94720-7450, USA}

\author{Ivan Vasko}
\affil{William B. Hanson Center for Space Sciences, University of Texas at Dallas, Richardson, TX, USA}
\author{Stuart D. Bale}
\affil{Space Sciences Laboratory, University of California, Berkeley, CA 94720-7450, USA}
\affil{Physics Department, University of California, Berkeley, CA 94720-7300, USA}

\author{Benjamin D.~G. {Chandran}}
\affil{Department of Physics \& Astronomy, University of New Hampshire, Durham, NH 03824, USA}
\author{{Alexandros Chasapis}}
\affil{Laboratory for Atmospheric and Space Physics, {University of Colorado}, {Boulder}, {80303}, {CO}, {USA}}
\author{Thierry {Dudok de Wit}}
\affil{LPC2E, CNRS and University of Orl\'eans, Orl\'eans, France}
\affil{ISSI, Bern, Switzerland}
\author{Alfred Mallet}
\affil{Space Sciences Laboratory, University of California, Berkeley, CA 94720-7450, USA}
\author{Michael McManus}
\affil{Space Sciences Laboratory, University of California, Berkeley, CA 94720-7450, USA}
\author{Romain Meyrand}
\affil{Department of Physics, University of Otago, 730 Cumberland St., Dunedin 9016, New
Zealand}

\author{Marc Pulupa}
\affil{Space Sciences Laboratory, University of California, Berkeley, CA 94720-7450, USA}
\author{Jonathan Squire}
\affil{Department of Physics, University of Otago, 730 Cumberland St., Dunedin 9016, New
Zealand}
\begin{abstract}
Circularly polarized, nearly parallel propagating waves are prevalent in the solar wind at ion-kinetic scales. At these scales, the spectrum of turbulent fluctuations in the solar wind steepens, often called the transition-range, before flattening at sub-ion scales. Circularly polarized waves have been proposed as a mechanism to couple electromagnetic fluctuations to ion gyromotion, enabling ion-scale dissipation that results in observed ion-scale steepening.   
Here, we study Parker Solar Probe observations of an extended stream of fast solar wind ranging from $\sim 15 R_\odot - 55 R_\odot$. We demonstrate that, throughout the stream, transition-range steepening at ion-scales is associated with the presence of significant left-handed ion-kinetic scale waves, which are thought to be ion-cyclotron waves. We implement quasilinear theory to compute the rate at which ions are heated via cyclotron resonance with the observed circularly polarized waves given the empirically measured proton velocity distribution functions. We apply the Von K\'arm\'an decay law to estimate the turbulent decay of the large-scale fluctuations, which is equal to the turbulent energy cascade rate. We find that the ion-cyclotron heating rates are correlated with, and amount to a significant fraction of, the turbulent energy cascade rate, implying that cyclotron heating is an important dissipation mechanism in the solar wind. 
\end{abstract}

\section{Introduction}

  Weakly collisional plasmas, which are common in astrophysical environments, are fundamentally governed by kinetic processes \citep{Marsch2006}. Our understanding of kinetic processes responsible for turbulent dissipation, heating, and energy transfer in collisionless environments is relatively incomplete, and necessary to explain phenomena such as solar wind acceleration and coronal heating \citep{Parker1958,Richardson1995,Hellinger2013,Fox2016}, and analogous astrophysical processes.

Recent work on the near-Sun solar wind has highlighted the significant presence of circularly polarized ion-scale waves \citep{Bale2019,Bowen2020a}, and their association with non-thermal features in particle distributions \citep{Verniero2020,Klein2021,Verniero2021}.  These waves are characterized by their quasi-parallel propagation along the  magnetic field \citep{Jian2014,Boardsen2015,Bowen2020a,Liu2023}. Electric field measurements suggest that these wave predominantly propagate outward from the sun \citep{Bowen2020d}, similar to the observed propagation direction of larger-scale Alfv\'{e}nic turbulent fluctuations \citep{Roberts1987a,TuMarsch1995,Bavassano1998,McManus2020}. 

While ion-scale waves are often associated with processes related to kinetic plasma distributions, e.g. instabilities and resonant damping, \citep{Gary1993,IsenbergLee1996,Hollweg2002,Marsch2006,Klein2018,Klein2021}, they may also serve as a mechanism to transfer turbulent energy via cyclotron resonance to particle thermal motion \citep{HollwegJohnson1988,TuMarsch1997,Cranmer2000,HollwegIsenberg2002,Cranmer2014}. 
Signatures of quasilinear cyclotron resonance in the observed proton distribution functions have been suggested in various spacecraft observations \citep{MarschTu2001b,He2015,Verniero2021,Bowen2022}. Furthermore, steepening of turbulent spectra at ion cyclotron resonant scales has been interpreted as a signature of cyclotron resonant damping \citep{Denskat1983,Woodham2018,Lotz2023} and is correlated with the presence of circularly polarized signatures \citep{Goldstein1994,Leamon1998a,He2011,Lion2016,Zhao2021}. These helical signatures are both correlated with proton temperature anisotropy as well as the turbulent amplitudes \citep{Telloni2019}, suggesting they may be associated with turbulent dissipation.

The quasilinear damping of the {\em{in situ}} population of ion cyclotron waves (ICW) has been measured with a heating rate accounting for 10-20\% of the turbulent energy flux \citep{Bowen2022}. Further studies have shown the direct transfer of energy from waves to protons using wave-particle correlation methods \citep{Vech2020, Luo2022}. The presence of these waves has been observed to correlate both to turbulent features such as the large-scale cross helicity and sub-ion-scale intermittency \citep{Bowen2023}. 

In this Letter, we study an extended stream of fast solar wind observed by PSP from $\sim15$ to $55 R_\odot$ with persistent signatures of left-hand polarized waves.{ We show that left-handed polarization directly corresponds to turbulent steepening in the ion-kinetic scale transition range \citep{Sahraoui2009,Kiyani2009,Bowen2020c,Duan2020,Duan2021}, which has historically been interpreted as a signature of dissipation \citep{Denskat1983,Goldstein1994,Leamon1998a,Smith2006,Smith2012,Lion2016,Bowen2020c,Bowen2023}}. Using the cold plasma dispersion to determine the internal energy and Poynting flux of the measured wave spectrum suggests that the energy to generate waves is stored within the turbulent fluctuations, indicating that the waves are a pathway to ion-scale turbulent dissipation. By applying drifting bi-Maxwellian fits to the distribution function, we estimate the empirical quasilinear heating rate of the local ion-scale waves \citep{KennelEngelmann1966,IsenbergLee1996,Bowen2022}. We similarly estimate local turbulent dissipation rates via turbulent amplitudes and the Von K\'{a}rm\'{a}n decay law to determine the decay of the largest, outer-scale turbulent fluctuations \citep{Hossain1995,Wan2012,Bandyopadhyay2020,Wu2022}. We demonstrate strong correlations between the quasilinear heating and turbulent dissipation rates. These strong correlations are evident in both global scaling as well as local fluctuations in the turbulent cascade and quasilinear heating rates. These results suggest that cyclotron heating plays an important role in extended solar wind heating. These results may have strong implications for the nature of ion-scale heating in the corona and other weakly-collisonal astrophysical plasmas.


\begin{figure}
        \includegraphics[width=\columnwidth]{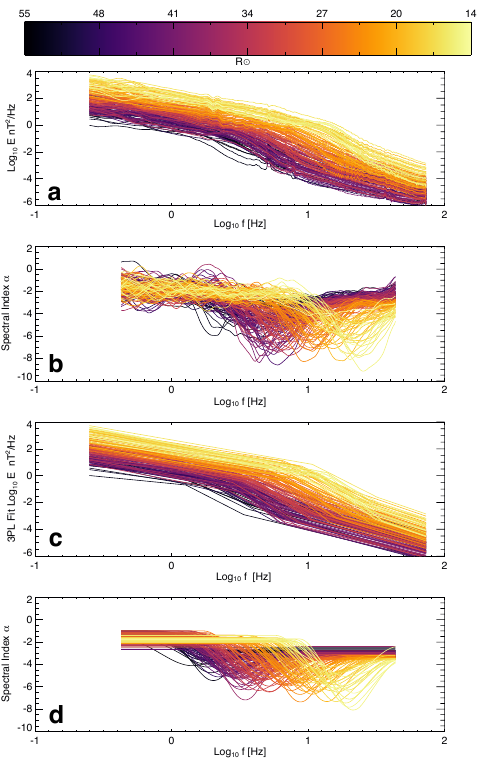}
    \caption{a) Measured spectral densities from Nov 16-20, 2021 computed for 6737 intervals of 128 s. Every 40th spectra is plotted. Colors represent radial distance of PSP from the sun, ranging from $\sim 14-55$ $R_\odot$, with light colors representing small distances becoming darker to indicate increasing distance. b) Local spectral index $\alpha$ of measured spectral densities computed in moving window. c) Three power-law fit (3PL) to observed spectra for each interval. d) Local spectral index $\alpha$ of 3PL spectral densities computed in moving window.}
    \label{fig1}
\end{figure}
\section{Methods \& Results}

\begin{figure}
    \centering
    \includegraphics[width=3.5in]{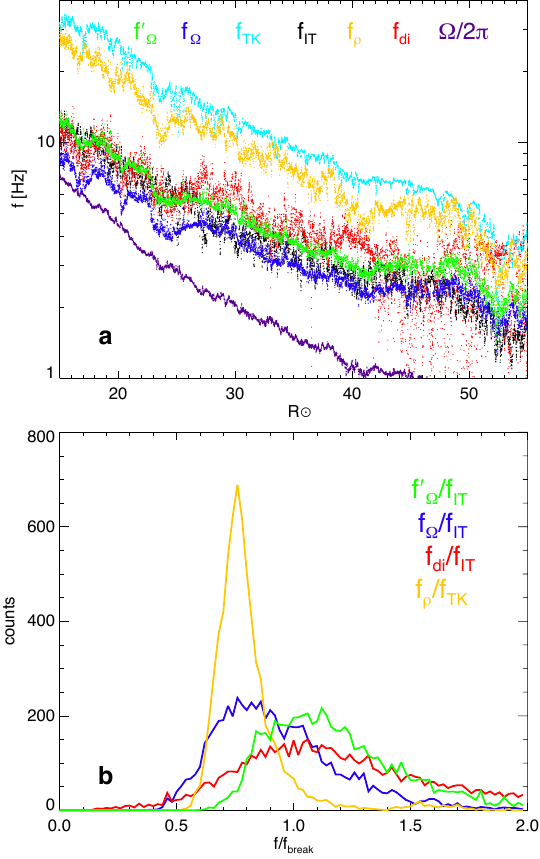}
    \caption{a) Frequencies corresponding to ion-kinetic scales computed with the Taylor Hypothesis for 6737 intervals as function of solar radius, ion-inertial scale $f_{d_i}$ (red), ion gyroscale $f_{\rho_i}$ (orange), cyclotron resonant scale $f_\Omega$ (blue). Additionally a correction to the Taylor Hypothesis approximation to the cyclotron scale $f'_\Omega$ (green). Break points from spectral fits are also shown $f_{IT}$ (black) and $f_{TK}$ (teal). The gyrofrequency $\Omega/2\pi$ is shown in purple. b) Histograms show the distribution of frequency ratios $f_{{d}_i}/f_{IT}$ (red), $f_{\Omega}/f_{IT}$ (blue), $f'_{\Omega}/f_{IT}$ (green) and  $f_{{\rho}_i}/f_{TK}$ (orange).}
    \label{fig2}
\end{figure}
\begin{figure*}
\centering
    \includegraphics[width=6.5in]{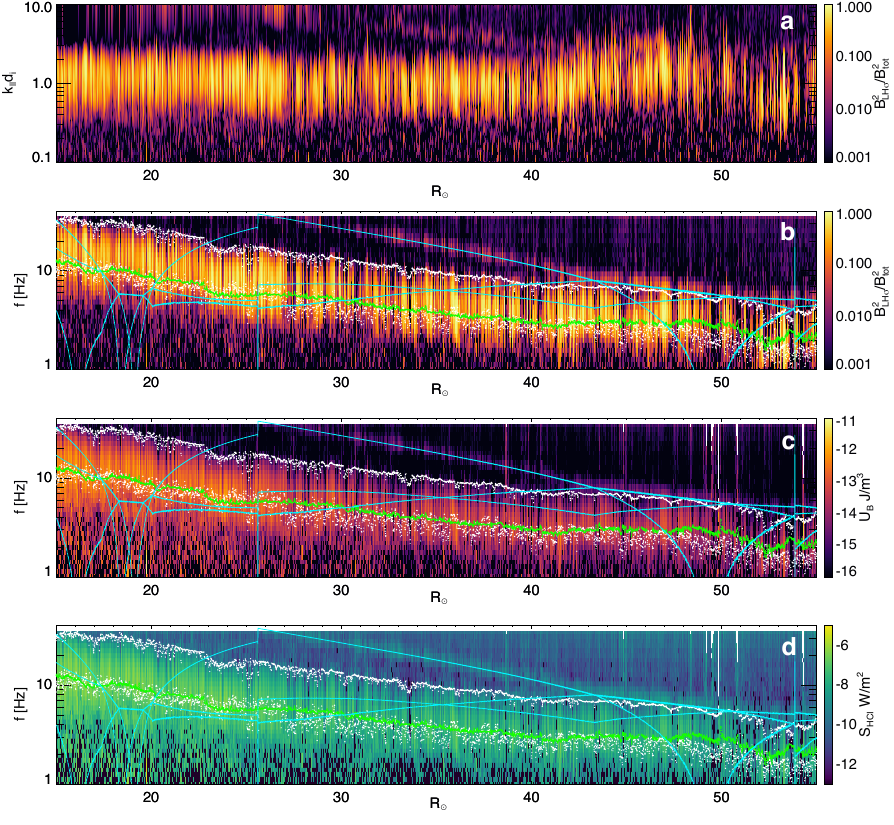}
   \caption{ a) Fractional left-handed power $\tilde{B}^2_{LH\sigma}/\tilde{B}^2_{total}$ computed from wavelet-transform for each of the 6737 intervals as a function of $k_\parallel d_i$. b) Spectrogram of left-handed circularly polarized power as function of radial distance; $f_{IT}$  and $f_{TK}$ for each measurement are plotted in white, as well as the cyclotron scale $f_\Omega$  in green; teal lines show frequencies corresponding to spacecraft reaction wheels \citep{Bowen2020a}. c) internal energy  $U_B$ and d) Poynting flux of waves in a heliocentric inertial frame $S_{\rm HCI}$ computed from the cold-plasma dispersion relation.}
    \label{fig3}
\end{figure*}

We focus on a stream of fast solar wind observed in the inner heliosphere by Parker Solar Probe (PSP) from 11/16/2021 to 11/20/2021. During this interval PSP was flying inwards towards the sun in near co-rotation with the solar surface, sampling a relatively singular source region over the range of 14-55 $R\odot$ \citep{Badman2023,Davis2023}.  We use measurements from the PSP Solar wind Electron Alpha and Proton experiment's Solar Probe Analyzer \citep[SPAN]{Livi2022}. The PSP FIELDS experiment \citep{Bale2016} provides measurements of the magnetic field from DC to sub-ion kinetic scales using merged search coil \citep{Jannet2021} and fluxgate magnetometer (SCaM) measurements \citep{Bowen2020b}. 

We break the 4 day stream into a set of intervals of 128 s with 50\% overlap to study the evolution of the turbulent spectra alongside ion-scale waves. SCaM data are sensitive well into the sub-ion kinetic scales \citep{DudokdeWit2022},  but are only available for two axes{; as a result, the SCaM data are only used to measure the shape of the spectra. The fluxgate magnetometer data provides three component magnetic field measurements that enable comprehensive study of the properties of observed waves.} The fluxgate magnetometer and SCaM data are re-sampled to a uniform 146.4845 Sa/sec rate. Data were discarded when discontinuities or changes in instrumental modes led to artifacts. In total 6737 intervals of 128 s were analyzed.

\paragraph{Turbulent Spectra}

We compute spectral densities $E(f)${, with units {nT\textsuperscript{2}/Hz},} from the two-component SCaM data through ensemble averaging over 4096 point FFTs in each 128s interval. Each spectral density is interpolated onto 320 logarithmically spaced frequencies. Figure \ref{fig1}(a) shows spectral densities computed from the strem intervals, though for clarity only every 40th spectra is plotted. The colors corresponding to radial distance from the sun, which is between 14.1-55.0 $R_\odot$; lighter yellow colors are closer to the Sun, whereas darker colors correspond to further heliocentric distances.  

An estimate of the local spectral index for each frequency is obtained by performing a linear least-square fit of $\text{log}_{10} E(f)$ in a moving window consisting of 60 neighboring logarithmically spaced frequencies. The slope of a linear least-square fit in log space provides a measure of the local spectral index $\alpha(f)$. Figure \ref{fig1}(b) shows the locally measured $\alpha$ as a function of frequency for spectra in Figure \ref{fig1}(a).  The observed spectra are consistent with three-power-law spectra often reported in the solar wind \citep{Sahraoui2009, Alexandrova2008,Bowen2020c}: each spectra has a characteristic low-frequency inertial range scaling $\alpha_I$, a transition range with steepened index $\alpha_T$, followed by flattening to an index $\alpha_K$ at higher frequencies. {The shape of the spectra is consistent but shifts to higher frequencies closer to the Sun. The index of the steep transition range spectra varies, and can have values of up to -10. The extremely steep transition range indices in this stream are predominantly associated with parallel spectra as shown in \cite{Duan2021}. As we demonstrate, this steepening is also associated with significant ICW populations \citep{Bowen2020a,Bowen2023}.}

{To determine the slope of the transition range $\alpha_T$ we  implement a piecewise, three power-law (3PL) fit, following methods developed in \cite{Bowen2020c}. Each of the three ranges is modeled as a power-law spectra with indices $\alpha_I$, $\alpha_T$, and $\alpha_K$ separated by break-point frequencies $f_{IT}$ and $f_{TK}$, referring to the respective inertial-transition and transition-kinetic breaks. Fig. 1(c) shows the 3PL fits, corresponding to the spectra in Figure \ref{fig1}(a). Additionally, we apply the procedure to measure the local spectral index $\alpha(f)$ to the 3PL-fits, with the results shown in Figure \ref{fig1}(d). There is good agreement between the observations and the 3PL fits.}


\paragraph{Break Scales \& Cyclotron Resonance}

\begin{figure}
    \includegraphics[width=\columnwidth]{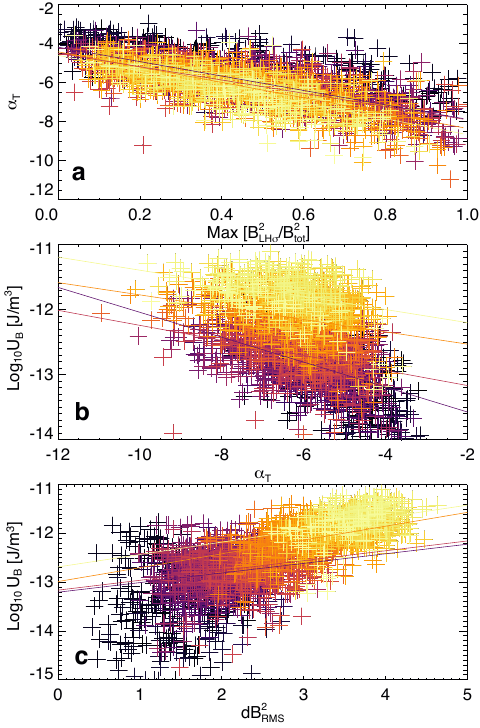}
    \caption{a) Transition range spectral index $\alpha_T$ against maximum circular polarization $\text{max} [\tilde{B}^2_{LH\sigma}/\tilde{B}^2_{total}]$. b) Total internal energy of cyclotron waves $U_B$ plotted against $\alpha_T$ . c) Total internal energy of cyclotron waves $U_{B}$, which is bandwidth limited and filtered by left hand polarization at ion-kinetic scales,  plotted against total rms amplitude of magnetic field $dB^2_{rms}$ computed at 128 seconds. Data are colored by $R_{\odot}$ with larger heliocentric distances in darker shades. Solid lines show the linear regression for $10-20 R_\odot$ (yellow), $20-30 R_\odot$ (orange), $30-40 R_\odot$ (red), and $40-50 R_\odot$ (purple), darker colors correspond to larger distances.}
    \label{fig4}
\end{figure}
We compare the measured break scales $f_{IT}$ and $f_{TK}$ to various physical ion-kinetic scales: the ion gyroradius,  $\rho_i = v_{thi}/\Omega_i$, where the ion thermal speed is $v_{thi}=\sqrt{2k_BT_{i}/m_i}$ and the ion gyrofrequency is $\Omega_i=eB_0/m_i$; and the ion inertial scale $ d_i=\rho_i/\sqrt{\beta_i}$, where $\beta_i=2n_0\mu_0T_i/B_0^2$. The fundamental charge is given as $e$, $\mu_0$ is the permeability of free space, $T_i$ is the proton temperature, $B_0$ is the average background magnetic field, $n_0$ is the background number density. The wavenumbers corresponding to the ion-inertial and ion gyroscale are $k_{\rho_i}={\rho_i}^{-1}$ and $k_{d_i}=d_i^{-1}$. We approximate the wavenumber corresponding to cyclotron resonant interactions as  $k_\Omega=\Omega_i/(v_A +v_{thi})$, where $v_A$ is the Alfv\'{e}n speed $v_A=B_0/\sqrt{mu_0m_in_0}$
\citep{Leamon1998a,Wicks2016,Woodham2018,Bowen2020a}. This approximation for $k_\Omega$ is derived via the cyclotron resonance condition between outward going ICW and protons flowing inwards in the plasma frame,  $\omega + {k}_\parallel v_{th}= \Omega_i$ \citep{Leamon1998a}, and a low frequency limit $\omega=k_\parallel v_A$ to the ion-cyclotron dispersion for parallel propagating waves. The cold-plasma dispersion relation is

\begin{equation}
\left(\frac{\omega_{ICW}}{\Omega_i}\right)^2 =\left[ \frac{k_\parallel d_i}{2}\left(
\sqrt{k_\parallel^2 d_i^2+4}- k_\parallel d_i\right)\right]^2 \label{eq:disp}.\end{equation}

For the ion-scales  $k_{d_i}$, $k_{\rho_i}$, and $k_\Omega$, each wavenumber is converted to an effective spacecraft frequency $f$ 
first using the Taylor hypothesis, $2\pi f\approx {k}{v}_{sw}$ to associate ion-kinetic scales with spacecraft frequencies $f_{{d}_i}$,$f_{{\rho}_i}$, and $f_\Omega$. For the cyclotron-resonant scale, $k_\Omega$, we also consider a correction to the Taylor hypothesis made by incorporating the Doppler shift equation
\begin{equation}
2\pi f =\omega(k) + \mathbf{k}\cdot {\mathbf{v}}_{sw}, \label{eq:dop}
\end{equation}
to  improve on our estimate of the frequency corresponding to cyclotron resonant interactions as
\begin{equation}
    2\pi f'_\Omega =\omega_{ICW}(k_\Omega) + {k_\Omega}{{v}}_{sw}, 
\end{equation}
which includes the contribution from $\omega_{ICW}$ evaluated at $k_\Omega$. For calculating $f'_\Omega$, we assume that the angle between the mean magnetic field and flow direction, $\theta_{BV}$, is small such that $cos\theta_{BV}\approx 1$ and thus $\mathbf{k}\cdot {\mathbf{v}}_{sw}\approx{k_\Omega}{{v}}_{sw}$ for parallel propagating ICWs.

Fig. \ref{fig2}(a) shows the computed kinetic-scale frequencies alongside the measured breaks from the 3PL fitting algorithm against $R_\odot$. 
Fig. \ref{fig2}(b) shows distributions ratios of $f_{{d}_i}/f_{IT}$ $f_\Omega/f_{IT}$, $f'_\Omega/f_{IT}$ and $f_{{\rho}_i}/f_{TK}$. These distributions provide a measure of agreement between break frequencies and the measured ion kinetic scales.
Recent work has suggested that the cyclotron scale $f_\Omega$ corresponds well with $f_{IT}$ \citep{Woodham2018,Vech2018,Duan2020,Lotz2023}, here we find that $f_\Omega$ is statistically lower than $f_{IT}$; however, the corrected $f'_\Omega$ frequency well approximates the $f_{IT}$ break. The $f_{IT}$ transition range break additionally approximates $f_{{d}_i}$, such that we cannot distinguish the inertial scale from the cyclotron resonant scale. We find that $f_{{\rho}_i}$ is within the transition range, which is consistent with previous results \citep{Bowen2020c}.



Importantly, $f_\Omega$ and $f'_\Omega$ are estimates to a single cyclotron resonant scale corresponding to resonance between particles at the thermal speed $v_{th}$,  with low-frequency ICWs, $\omega =kv_A $. However, observations from the solar wind suggest that ICWs occur over a range of frequencies corresponding to $kd_i\approx1$ \citep{Bowen2020a}. Empirical determination of circular polarization as a function of spacecraft frequency along with the cold plasma dispersion for parallel propagating ICWs Eq. \eqref{eq:disp}
provides information regarding the wavenumbers at which cyclotron resonant waves occur \citep{Bowen2020d}.

A 64-scale Morlet wavelet transform
is applied to each interval to compute a power spectral density $\tilde{B}^2 (f,t)$ with units nT$^2$/Hz \citep{Farge1992,DudokdeWit2013,Bowen2020a}. We extract circular polarization of the field via the {reduced} magnetic-helicity \citep{HowesQuataert2010},
\begin{align}
        \sigma_B(f,t)=-2\text{Im}(\tilde{B}_{\perp1}\tilde{B}_{\perp2}^*)/(\tilde{B}_{\perp1}^2+\tilde{B}_{\perp2}^2).
\end{align}
of the wavelet coefficients perpendicular to the mean field with left/right handed polarization represented by positive/negative $\sigma_B$.  We calculate the left-handed polarized power spectra $\tilde{B}^2_{LH\sigma}$ by filtering out power with $\sigma_B<0.9$. The normalized fractional left handed spectra are computed as $\tilde{B}^2_{LH\sigma}/\tilde{B}^2_{total}$. Very little right handed polarization is present in this stream.

Based on previous measurements of the phase speed of circularly polarized waves \citep{Bowen2020d}, which show a strong statistical preference for outward-propagation, we can assume that the waves are outward-propagating ICWs, travelling parallel to the mean field. The wavelet transform gives circular polarization as a function of the spacecraft frame frequency, $f$. The use of the Doppler shift Eq.~\eqref{eq:dop} in the parallel-propagating limit gives
 \begin{equation}
     2\pi f =\omega_{ICW} \pm k_\parallel v_{sw} cos\theta_{BV} \label{eq:pardop},\end{equation}
     
     where we no longer assume either the Taylor Hypothesis or that $cos\theta_{BV}\approx 1$. The combination of Eq. \eqref{eq:pardop} with Eq. \eqref{eq:disp} determines the corresponding parallel wave-number $k_\parallel$ for each spacecraft frequency $f$ \citep{Bowen2020d}. Figure \ref{fig3}(a) shows the normalized $\tilde{B}^2_{LH\sigma}/\tilde{B}^2_{total}$ as function of $ k_\parallel d_i$ computed from combining Eq. \eqref{eq:disp} with Eq. \eqref{eq:pardop}. At all radial distances the ICW population appears at $k_\parallel d_i=1$, where ICWs are inherently dispersive, with a finite bandwidth in wave number ranging from approximately $k_\parallel d_i\sim 0.3-4$. The finite bandwidth indicates that, at each time, a range of particle velocities simultaneously satisfy the ICW resonance condition. It is important to note that the spacecraft reaction wheels, highlighted in teal lines in Figure \ref{fig3}(b-d), contribute circularly polarized power at higher $k_\parallel d_i$ \citep{Bowen2020a}.

 Fig. \ref{fig3}(b) shows $\tilde{B}^2_{LH\sigma}/\tilde{B}^2_{total}$ as a function of spacecraft frequency at each interval studied as a function of solar radial distance. We plot both the $f_{IT}$ and $f_{TK}$ breaks. The $f_{IT}$ break agrees very well with the circularly polarized regime, while the $f_{TK}$ break bounds the circular polarization at higher frequencies, indicating that the transition range is entirely circularly polarized. We additionally plot the cyclotron scale $f'_\Omega$, which approximates the $f_{IT}$ break.

\paragraph{Cyclotron Waves and Turbulent Dissipation}

Using the cold-plasma dispersion and the Doppler shift Eqs.~\eqref{eq:disp} \& \eqref{eq:dop} and our observations of $\tilde{B}^2_{LH\sigma}(k_\parallel d_i)$ we measure both the internal energy density of the waves $U_B$ and the magnitude of the Poynting flux computed in an Heliocentric inertial frame (HCI), $S_{\rm HCI}$ assuming the ICW propagation is entirely parallel the radial solar wind flow \citep{Karpman1974,Shklyar}:

\begin{align}
    U'_B=\frac{\tilde{B}^2_{LH\sigma}}{\mu_0}\frac{v_{ph}}{v_g}=\frac{\tilde{B}^2_{LH\sigma}}{2\mu_0}\frac{(2\Omega_i-\omega)}{(\Omega_i-\omega)}
,\\
  S_{\rm HCI}'=(v_{sw}+v_{g})U_B',
\end{align}
where $v_{ph}$ and $v_g$ are the plasma frame phase and group velocities respectively. As the units of ${\tilde{B}^2_{LH\sigma}}$ are nT$^2$/Hz both $U'_B$ and $S_{\rm HCI}'$ are defined as spectral densities. The internal energy $U_B$ and Poynting flux magnitude $S_{\rm HCI}$ corresponding to each wavelet coefficient are obtained by multiplying $U'_B$ and $S_{\rm HCI}'$ by the bandwidth of each wavelet $\Delta f= \frac{df}{dk_\parallel}\Delta k_\parallel$, where the derivative ${df}/{dk_\parallel}$ and $\Delta k_\parallel$ are obtained from Eq. \ref{eq:pardop}. Figure \ref{fig3} (c\&d) show $U_B$ and $S_{\rm HCI}$ at each frequency over the interval.


The total Poynting flux and internal energy can be integrated at each time to give total values  $U_{tot}$ and $S_{tot}$ for each interval. Due to the sporadic noise that occurs at low frequencies, as well as high frequency contributions to polarization from the spacecraft reaction wheels, we zero out contributions to $S_{tot}$ and $U_{tot}$ where $kd_i<0.3$ and $kd_i>5$.


We use $U_{tot}$ and $B^{\text{max}}_{LH\sigma}=\text{max} [\tilde{B}^2_{LH\sigma}/\tilde{B}^2_{total}]$  as proxies for the ICWs in each interval.  Fig. \ref{fig4} shows how these quantities relate to parameters associated with turbulence, $\alpha_T$ and as well as the total RMS turbulent amplitude $\langle\delta B^2\rangle$ computed in each 128 s interval.  We consider only intervals with $150^\circ<\theta_{BV}<170^\circ$ in order to control for effects associated with anisotropy, which affect observational signatures of both the waves and the turbulence \citep{Chen2010a,Bowen2020a}; of the total 6736 intervals, 3172 occur with $150^\circ<\theta_{BV}<170^\circ$. In each panel of Fig.~\ref{fig4} data is plotted with the color scale in Fig.~\ref{fig1} with lighter colors corresponding to intervals closer to the sun. We compute lines of best fit for data within $10-20 R_\odot$, $20-30 R_\odot$, $30-40 R_\odot$, and $40-50 R_\odot$, which are shown in each panel of Fig. \ref{fig4}.  Fig. \ref{fig4}(a) shows that the transition range slope $\alpha_T$ is strongly correlated with the level of circular polarization, $B^{\text{max}}_{LH\sigma}$ \citep{Bowen2023}, with similar trends observed at all solar radii. Fig. \ref{fig4}(b) shows that the internal energy of the waves $U_B$ is anti-correlated with $\alpha_T$, such that steeper slopes contain greater amounts of ICW energy. This correlation is present at all $R_\odot$, though at larger distances the maximum measured $U_{tot}$ are similar to the lowest $U_{tot}$ close to the sun, such that constraining the data by $R\odot$ is necessary to see the correlations. Fig. \ref{fig4}(c) shows that the internal energy contained within the circularly polarized ion-scale transition range is globally a function of the amplitude of the turbulent fluctuations $dB_{rms}^2$ computed at the 128 s scale, consistent with previous results in \cite{Shankarappa2023}. 
\paragraph{ICW Heating Rates}

Our analysis of the transition range reveals signatures of circularly polarized ion-cyclotron resonant waves; we apply well established principles of energy conservation and quasilinear heating to connect these waves to turbulent dissipation. 

The Poynting theorem states that the change in internal energy of the ICW population is
\begin{equation}
    \frac{\partial U_B}{\partial t}=-\nabla\cdot \mathbf{S} -Q_{\rm ICW}+\epsilon_{\rm ICW},
    \label{eq:poynt}
\end{equation}
where $Q_{\rm ICW}$ is the dissipation rate of the waves and $\epsilon_{\rm ICW}$ is a driving term. Assuming a steady state, we compute the three terms on the right side of Eq. \eqref{eq:poynt}.
We approximate $\nabla\cdot \mathbf{S}$ in the heliocentric inertial frame as the radial derivative

\begin{equation}\nabla\cdot \mathbf{S}\approx \frac{1}{r^2}\frac{\partial}{\partial r}r^2 S_{tot}.\end{equation}
We estimate $Q_{\rm ICW}$ using the quasilinear heating rate \citep{KennelEngelmann1966} with the empirically observed spectrum of cyclotron waves \begin{align}
I(k_\parallel)=\frac{\tilde{B}^2_{LH\sigma}}{B_0^2 }\frac{df}{dk_\parallel}.\end{align}  For each SPANi measurement, we perform a drifting bi-Maxwellian fit in the proton core frame with the form \begin{align}
\begin{split}
g_p(v_\perp,v_\parallel)=& \frac{n_c}{\pi^{3/2}w_{c,\perp}^2 w_{c,\parallel}}\text{exp}\left[-\frac{v_\perp^2}{w_{c,\perp}^2} -\frac{v_\parallel^2}{w_{c,\parallel}^2}\right]\\
&+\frac{n_b}{\pi^{3/2}w_{b,\perp}^2 w_{b,\parallel}}\text{exp}\left[-\frac{v_\perp^2}{w_{b,\perp}^2} -\frac{(v_\parallel-v_D)^2}{w_{b,\parallel}^2}\right],
\label{eq:bimax}
\end{split}
\end{align}
which has anisotropic thermal speeds perpendicular and parallel the mean field ($\perp$ and $\parallel)$ 
for the beam and core (subscripts, b and c): 
$w_{c,\perp},w_{c,\parallel}, w_{b,\perp},w_{b,\parallel}$ and a relative drift, $v_D$, parallel to the mean magnetic field \citep{Marsch2006,Klein2021}. We compute the average fit parameters over the 128s interval such that an average gyrotropic $\bar{g}(v_\perp,v_\parallel)$ is computed. Using $\bar{g}(v_\perp,v_\parallel)$, and the observed spectrum of cyclotron waves $I(k_\parallel)$, we use the cold plasma dispersion, Eq.~\ref{eq:disp}, with wave numbers defined from Eq.~\ref{eq:pardop}, to determine the volumetric heating rate in the local interval:

\begin{align}\label{eq:H}
 Q_{\rm ICW}=\int \frac{m_pv^2}{2}\frac{\partial\bar{g}}{\partial t} d^3\mathbf{v}& \nonumber\\ \nonumber=
 \frac{ \pi e^2}{4m_p}\int d^3\mathbf{v}\Bigg\{v^2 \int_0^\infty  dk_\parallel\frac{1}{v_\perp}&\hat{G}_k\Bigg[ v_\perp \delta(\omega_k -k_\parallel v_\parallel-\Omega_p)\\&\times\frac{\omega_k^2}{k_\parallel^2c^2} I(k_\parallel)\hat{G}_k \bar{g}(v_\perp,v_\parallel)\Bigg]\Bigg\}, 
\end{align}
    with
    \begin{align}
    \hat{G}_k=  (1- \frac{k_\parallel v_\parallel}{\omega_k})\frac{\partial}{\partial{v_\perp}} +\frac{k_\parallel v_\perp}{\omega_k}\frac{\partial}{\partial{v_\parallel}}\end{align} 
\citep{KennelEngelmann1966}. A gyrotropic differential volumetric heating rate $Q(v_\perp,v_\parallel)$ can be defined from $Q_{\rm ICW}=\int Q(v_\perp,v_\parallel)dv_\parallel dv_\perp$. We outline how to calculate $Q_{\rm ICW}$ and $Q(v_\perp,v_\parallel)$ in Appendix \ref{AppendA} 
through using the $\delta$-function to evaluate the integral over wave number and the subsequent numerical integration of $Q(v_\perp,v_\parallel)$ over velocity coordinates.

\begin{figure*}
    \centering
   \includegraphics[width=\textwidth]{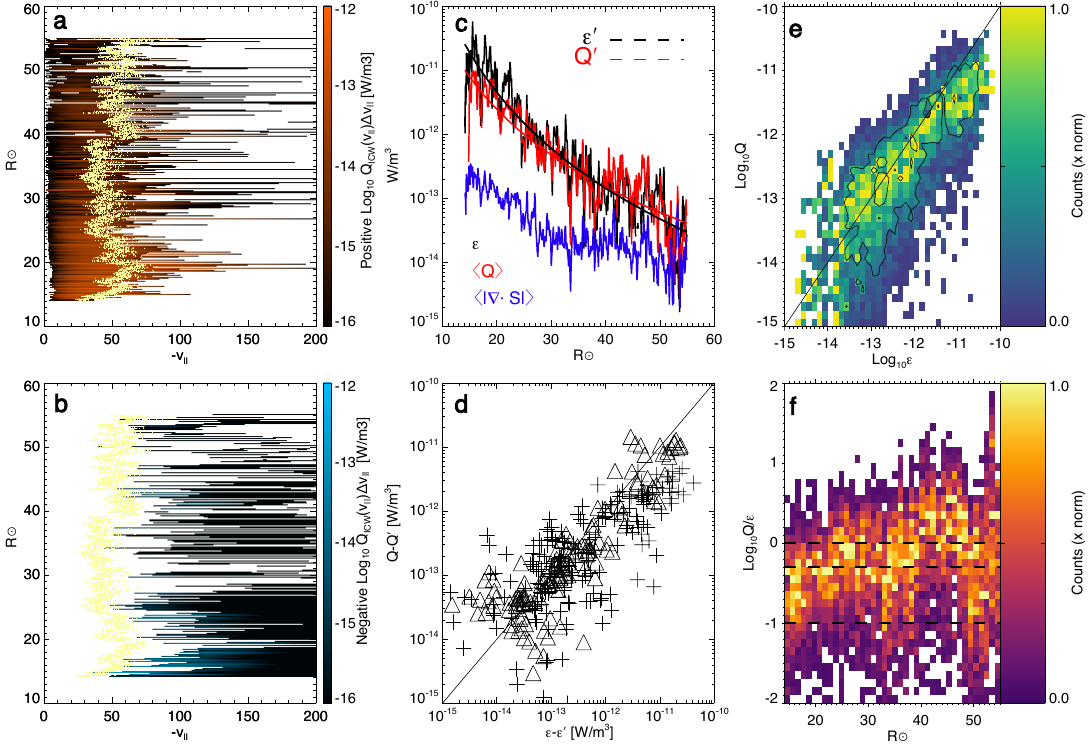}
   \caption{ a) Volumetric heating rates, $Q(v_\parallel)\Delta v_\parallel$, with $\Delta v_\parallel= 1\textrm{km/s}$, which have been integrated over perpendicular velocity, as a function of resonant parallel velocity and solar radial distance. b) Same as panel a, but for negative values of  $Q(v_\parallel)\Delta v_\parallel$.  Yellow dots in panels a\& b show the measured parallel thermal speed. c) Measured turbulent cascade rate $\epsilon$ from von K\'{a}rm\'{a}n decay law (black), average cyclotron heating rate from quasilinear theory $\langle Q_{\rm ICW}\rangle$ (red), and $\langle\nabla\cdot \mathbf{S}\rangle$ for total cyclotron waves (blue). Power-law fits, $\epsilon$' and $ Q_{\rm ICW}$', are shown in dashed lines. d)  $\langle Q_{\rm ICW}\rangle$ -$ Q_{\rm ICW}'$ plotted against $\epsilon-\epsilon'$, where + and $\triangle$ corresponds to data with $\langle Q_{\rm ICW}\rangle$ -$ Q_{\rm ICW}'$ and $\epsilon-\epsilon'$ respectively greater and less than zero. e) Two-dimensional histogram of $Q_{\rm ICW}$ against $\epsilon$ interpolated onto 128 second cadence. f) Two-dimensional histogram of $Q_{\rm ICW}/\epsilon$ against $R_\odot$; dashed black lines show $Q_{\rm ICW}/\epsilon$ at levels of 0.1, 0.5, and 1; the average value of $Q_{\rm ICW}/\epsilon$ is shown in black; $Q_{\rm ICW}/\epsilon >1 $ indicating quasilinear heating at levels greater than the turbulent cascade may arise due to uncertainty in the estimates of both $Q_{\rm ICW}$ and $\epsilon$. The distributions in panels e\&f are column normalized to the maximum value in each data column.}
    \label{fig5}
\end{figure*}

Figure \ref{fig5}(a-b) shows $Q(v_\parallel)$, the volumetric heating rate as function of resonant parallel velocity, which is computed from integrating $Q(v_\perp,v_\parallel)$ over perpendicular velocities, as a function of solar radius. 
Figure \ref{fig5}(a) shows positive $Q(v_\parallel)\Delta v_\parallel$, at resonant velocities less than the thermal speed (shown in yellow), corresponding to wave-absorption and consequent plasma heating. The $\Delta v_\parallel$ term, which is the resolution of the numerical integration 1 km/s,
is included to normalize the differential $Q(v_\parallel)$ in terms of a volumetric heating rate W/m$^3$.  Figure \ref{fig5}(b) shows wave emission, which, consistent with \cite{Bowen2022}, is significantly less energetically relevant than the regions with positive $Q(v_\parallel)\Delta v_\parallel$ wave-absorption, which indicates net resonant-heating of the plasma. 

Importantly, the use of a biMaxwellian model to approximate the distribution may not capture non-thermal features that resonate with the observed populations of waves \citep{Dum1980,Vinas2009}. In Appendix \ref{AppendB}, we implement two well understood non-parametric techniques to model the observed distributions, Hermite polynomials and radial basis functions, to verify the independence of our results from the biMaxwellian model. While there are slight differences in the level of heating predicted by each model, the qualitative interpretation of ICW resonance heating is found in each case and the average values closely follow that of the drifting biMaxwellian.

\paragraph{Connecting ICW Heating to the Turbulent Cascade} The importance of resonant heating can be understood by comparing the net heating rates $Q_{\rm ICW}$ to the turbulent cascade rate $\epsilon$. Following \citep{Bandyopadhyay2020} we compute the energy cascade rate of the turbulence assuming a von K\'{a}rm\'{a}n decay law. 
\begin{equation}
   \epsilon'^{\pm} = \alpha \frac{(\delta z^\pm)^2 \delta z^\mp }{L^\pm}, \label{eq:VK}
\end{equation} where $\delta z^\pm$ are root mean square fluctuation amplitudes of the Elsasser variables $\delta\mathbf{z}^\pm= \delta\mathbf{v} \pm \delta\mathbf{B}/\sqrt{\mu_0\rho}.$ The correlation length $L^{\pm}$ is determined from the solar-wind advected spatial-scale corresponding to the time-lag for which the correlation reduces below a factor $e^{-1}$. The coefficient $\alpha=0.03$ is determined from numerical means \citep{Hossain1995,Wan2012,Usmanov2014,Bandyopadhyay2018} and chosen to agree with \cite{Bandyopadhyay2020} and \cite{Wu2022}. However, this value of $\alpha$ is derived assuming low cross helicity, while the near-Sun solar wind is characterized by imbalance with  $\sigma_c\approx 1$ \citep{McManus2020}. Note the units for $\epsilon'$ in equation \ref{eq:VK} are of J/s/kg; we define $\epsilon$ with units of W/m$^3$, which are comparable to the volumetric heating rates in Equation \ref{eq:H}, through multiplying $\epsilon'$ by the background mass density $\epsilon=\epsilon'm_in_0.$

Measurement of $\epsilon'^+$ via Eq. \eqref{eq:VK}, which we assume approximates the total cascade rate $\epsilon'$, requires the Elsasser amplitude $\delta z^{\pm}$ at the outer, energy containing scales, of 
 the turbulence. \cite{Davis2023} have studied the evolution of the outer scales in this same stream and found that the outer-scale spectral break evolves from $10^{-2}$ Hz close to 14 $R_\odot$ to $10^{-3}$ Hz at $\sim 50 R_{\odot}$. Accordingly, we compute the cascade rate $\epsilon^+$ in successive 30 minute intervals (corresponding to $\approx 5 \times 10^{-4}$ Hz) with $50\%$ overlap. In each 30 minute interval we then compute the average heating rate $\langle Q_{\rm ICW}\rangle$ and $\langle\nabla\cdot \mathbf{S}\rangle$. 

 Figure \ref{fig5}(c) shows $\epsilon$, $\langle Q_{\rm ICW}\rangle$ and $\langle\nabla\cdot \mathbf{S}\rangle$ as a function of radial distance. A surprisingly good agreement is found between $\epsilon$ and $\langle Q_{\rm ICW}\rangle$, while the divergence of the Poynting flux term is significantly smaller. The relatively small $\langle\nabla\cdot \mathbf{S}\rangle$ indicates that the wave dynamics are dominated by the source, $\epsilon_{\rm ICW}$, and dissipation, $Q_{\rm ICW}$,  and that the waves do not propagate significantly. This result is consistent with previous studies arguing for in situ driving of these waves \citep{Bowen2020a,Vech2020,Liu2023}.

 We find a Spearman, non-parametric, ranked correlation between $\epsilon$ and $\langle Q_{\rm ICW}\rangle$ of 0.79; we further compute the Pearson-correlation of the logarithmic quantities as 0.88. Both of these values indicate strong correlations. Furthermore, we fit each of $\epsilon$ and $\langle Q_{\rm ICW}\rangle$ to a power-law in $R_\odot$ noted as $\epsilon'$ and $ Q_{\rm ICW}'$. We compute the local variations from the global power-law trends using $\epsilon -\epsilon'$ and  $\langle Q_{\rm ICW}\rangle$ -$\ Q_{\rm ICW}'$ as a measure of how closely the local quasilinear heating rate follows the turbulent energy cascade rate. Figure \ref{fig5}(d) shows the local fluctuations $\langle Q_{\rm ICW}\rangle$ -$ Q_{\rm ICW}'$ against $\epsilon -\epsilon'$. We measure a Spearman ranked-correlation of 0.56.

Additionally, we interpolate $\epsilon$, measured at a 30 minute cadence, onto the 128 second cadence of the  $Q_{\rm ICW}$ measurements. Figure \ref{fig5}(e) shows a 2D-histogram of  $Q_{\rm ICW}$ against the interpolated $\epsilon$; the distribution is column-normalized to the measured cascade rate. Good correlations are obtained, with a Spearman ranked-correlation of 0.58. These results suggest that characteristically, the quasilinear heating rate is proportional to the energy cascade rate with a constant of proportionality near unity. Figure \ref{fig5}(f) shows a 2D-histogram of  $Q_{\rm ICW}/\epsilon$, where the interpolated  $\epsilon$ is again used, against solar radius $R_\odot$; the distribution is column-normalized to $R_\odot$. Implicitly, this suggests that much of the turbulent cascade flux enters the ICW population such that $Q_{\rm ICW}=\chi\epsilon^+$ with $\chi \approx 0.1-1$ over the vast majority of the observed stream. There is some radial trend, with $0.1<\chi <0.5$ at $\approx15-20 R_\odot$ and $\chi\approx 1$ at larger distances.  Values with $\chi >1$ suggest dissipation in excess of the turbulent cascade rate, we believe this is predominantly due to uncertainty in the measurements of $Q_{\rm ICW}$ and $\epsilon.$


\section{Conclusions}
The nature of the steepening of turbulent energy spectrum at ion-kinetic
scales has long been debated and is an important signature in understanding collisionless turbulent dissipation \citep{Denskat1983,Smith1990,Goldstein1994,Leamon1998a}. Studies of the location of the ion-kinetic scale break frequency suggest that the most likely candidate scale corresponds to resonance between outward going ICWs and the thermal ion population \citep{Woodham2018,Vech2018,Duan2020,Lotz2023}. Previous observations from 1 AU have suggested that ICW play a role in dissipating solar wind turbulence \citep{Leamon1998a,Lion2016,Woodham2019,Telloni2019,Zhao2021,Luo2022}. In this Letter, we concretely demonstrate that ion-scale spectral steepening is associated with circular polarization. The break between the inerital and transition ranges follows the regime of ICWs, and the spectral break between the transition range and subion-scale turbulence bounds the circularly polarized waves exactly at higher frequencies. We suggest that the transition range corresponds to cyclotron waves that are associated with turbulent dissipation and ion-scale heating \citep{Bowen2023}. Our results show that internal energy and momentum of the waves is correlated with the amplitude of the turbulent fluctuations, Figure \ref{fig3}, suggesting that the energy contained in, and transported by, the waves originates from the turbulent fluctuations. These results are consistent with earlier observations from 1 AU suggesting that the level of steepening in the turbulent cascade may relate to the level of Alfv\'{e}nicity and the turbulent cascade rate \citep{Smith2006,Bruno2014}.

In this Letter, we compute two measures of energy transfer: first, the quasilinear heating rate of kinetic scale ICWs, $\langle{Q_{\rm ICW}}\rangle$, and second, the von K\'{a}rm\'{a}n turbulent decay rate $\epsilon^+$. The measurements of $\langle{Q_{\rm ICW}}\rangle$ are obtained through integrating kinetic phase space densities measured by SPANi, while $\epsilon^+$ is determined from outer scale turbulent fluctuations. It is striking that these quantitative estimates of energy transfer, $\epsilon^+$ and ${Q_{\rm ICW}}$, obtained via entirely different methods at vastly different spatial scales show significant correlations. Such strong correlations suggest that parallel-cyclotron resonance dissipates significant amounts of turbulent energy into the solar-wind ion populations. The correlation between energy flux and quasilinear heating rate is observed in both the global scaling of the quantities as well as in local variations: i.e., fluctuations in the turbulent cascade rate are typically accompanied by correlated variations in the quasilinear heating rate. These novel measurements, which support a radially extended cyclotron heating mechanism, are important in understanding how turbulent dissipation results in heating the expanding solar wind. These results further indicate significant progress on PSP's objective ``to trace the flow of energy that heats and accelerates the solar corona and solar wind'' \citep{Fox2016}. Furthermore, the extended ICW heating, which is observed from 15 to 55 $R_\odot$, may have significant implications for the role of ICW heating in the corona, \citep{HollwegJohnson1988,Cranmer2000}.

Recent simulations by \cite{Squire2022} suggest that the dissipation of strongly imbalanced Alfv\'{e}nic turbulence may result in polarization signatures similar to those measured in the solar wind \citep{Podesta2011,He2011,Huang2020}. 
The \cite{Squire2022} simulations, and the underlying idea of a helicity barrier whereby imbalanced turbulence is prevented from cascading to small scales due to a conserved generalized helicity \citep{Meyrand2021}, provides a novel framework to understand the connection between turbulence and ion-cyclotron waves. These theoretical ideas have found support in observations showing that ion-scale ICWs preferentially occur when large-scale fluctuations are highly Alfv\'{e}nic \citep{Bowen2023} and that sub-ion scale intermittency depends largely on the level of ICWs present, which is in turn correlated to the cross-helicity, suggesting that ICWs play a role in the dissipation of imbalanced turbulence.

The idea that parallel propagating ICWs dissipate a predominantly perpendicular cascade challenges our current understandings of solar wind turbulence and dissipation. It has long been suggested that the turbulence should drive fluctuations into smaller perpendicular scales \citep{Shebalin1983}, and observations regularly indicate significant anisotropy perpendicular the mean field \citep{Horbury2008,Chen2010a,Duan2021}.  The hybrid simulations of the helicity barrier by \cite{Squire2022} suggest that oblique ICWs are the primary heating mechanism and that parallel ICWs are emitted as a secondary process \citep{Chandran2010b} through the Alfv\'{e}n/ion cyclotron instability \citep{Gary1993}.  

While the secondary emission of parallel ICWs from oblique-ICW heating provides a mechanism to generate parallel ICWs from turbulent heating, in such a mechanism, the plasma is heated by oblique ICWs but cooled by the parallel ICWs (since they are emitted as an instability). In contrast, our measurements indicate that the parallel ICWs are robustly heating the plasma, suggesting they should be directly driven by the turbulence, as opposed to generated via oblique ICW heating. If our present measurements, consistent with previous observations \citep{Bowen2022}, are correct, this presents a conundrum, given the difficulty of sourcing quasi-parallel waves from perpendicular turbulent structures.  Our use of non-parametric representations of $g(\mathbf{v})$ (Appendix \ref{AppendB}) further complicates this issue, as our measurement of ICW heating is not simply due to the incorrect parameterization of the SPANi observations with a drifting-bi-Maxwellian fit.

A possible source of systematic error is in the assumption of a cold-plasma. We assume a cold plasma dispersion \citep{Stix1992}, that is likely not entirely accurate for the solar wind. Implementation of warm-plasma dispersion solvers with arbitrary distribution functions \citep{Verscharen2017,Walters2023} may enable greater understanding of parallel wave-generation. Furthermore, inclusion of beam populations \citep{Verniero2020,Ofman2022} and $\alpha$-particles may affect instabilities \citep{McManus2023} and may help further understanding of these dynamics. 

Outside of uncertainties in the distribution function and wave-dispersion relations, a possible explanation for the in situ production of ICWs that result in solar wind heating could include the adiabatic evolution associated with expansion. Adiabatic evolution associated with expansion \citep{Chew1956} should drive distribution functions that are unstable and emitting ICWs towards stability. The extent that expansion may enable ICW heating rather than cooling within the helicity-barrier framework, remains largely unstudied.


Furthermore, estimates of the cascade rate are notoriously difficult to estimate via single spacecraft measurements \citep{Bandyopadhyay2018,Bandyopadhyay2020}. Care must be taken in understanding these cascade rates, especially in strongly imbalanced states, in which a stationary energy flux may not exist, or a negative cascade rate may occur \citep{Smith2009,Meyrand2021}. This is especially important in the regime of imbalanced turbulence frequently observed by PSP \citep{Meyrand2021,Squire2022}. However, recently \cite{Wu2022} found good agreement between the Von K\'{a}rm\'{a}n decay rates with measured perpendicular proton heating rates, suggesting that the turbulent dissipation can be studied via these methods. While care has to be taken in further studies of cascade rates, the general correspondence between the measured heating rates and the von Kárm\'{a}n decay rates obtained in this present work and by \cite{Wu2022} is promising.

In any case, these robust observations of extended cyclotron resonant process in the inner-heliosphere suggest that kinetic scale waves are strongly coupled to the turbulent cascade and provide an important pathway to the dissipation of turbulence in the solar wind and corona.

\section{Acknowledgements}
T.A.B. acknowledges NASA Grant No. 80NSSC24K0272. The work of I.V. was supported by NASA grant No. 80NSSC22K1634. B.D.G.C. acknowledges the support of NASA grant  80NSSC24K0171.
\bibliography{bib.bib}

\providecommand{\noopsort}[1]{}\providecommand{\singleletter}[1]{#1}%
\begin{thebibliography}{}
\expandafter\ifx\csname natexlab\endcsname\relax\def\natexlab#1{#1}\fi
\providecommand{\url}[1]{\href{#1}{#1}}
\providecommand{\dodoi}[1]{doi:~\href{http://doi.org/#1}{\nolinkurl{#1}}}
\providecommand{\doeprint}[1]{\href{http://ascl.net/#1}{\nolinkurl{http://ascl.net/#1}}}
\providecommand{\doarXiv}[1]{\href{https://arxiv.org/abs/#1}{\nolinkurl{https://arxiv.org/abs/#1}}}

\bibitem[{{Alexandrova} {et~al.}(2008){Alexandrova}, {Carbone}, {Veltri}, \&
  {Sorriso-Valvo}}]{Alexandrova2008}
{Alexandrova}, O., {Carbone}, V., {Veltri}, P., \& {Sorriso-Valvo}, L. 2008,
  The Astrophysical Journal, 674, 1153, \dodoi{10.1086/524056}

\bibitem[{{Badman} {et~al.}(2023){Badman}, {Riley}, {Jones}, {Kim}, {Allen},
  {Arge}, {Bale}, {Henney}, {Kasper}, {Mostafavi}, {Pogorelov}, {Raouafi},
  {Stevens}, \& {Verniero}}]{Badman2023}
{Badman}, S.~T., {Riley}, P., {Jones}, S.~I., {et~al.} 2023, arXiv e-prints,
  arXiv:2303.04852, \dodoi{10.48550/arXiv.2303.04852}

\bibitem[{Bale {et~al.}(2019)Bale, Badman, Bonnell, Bowen, Burgess, Case,
  Cattell, Chandran, Chaston, Chen, {et~al.}}]{Bale2019}
Bale, S., Badman, S., Bonnell, J., {et~al.} 2019, Nature, 1

\bibitem[{{Bale} {et~al.}(2016){Bale}, {Goetz}, {Harvey}, {Turin}, {Bonnell},
  {Dudok de Wit}, {Ergun}, {MacDowall}, {Pulupa}, {Andre}, {Bolton},
  {Bougeret}, {Bowen}, {Burgess}, {Cattell}, {Chandran}, {Chaston}, {Chen},
  {Choi}, {Connerney}, {Cranmer}, {Diaz-Aguado}, {Donakowski}, {Drake},
  {Farrell}, {Fergeau}, {Fermin}, {Fischer}, {Fox}, {Glaser}, {Goldstein},
  {Gordon}, {Hanson}, {Harris}, {Hayes}, {Hinze}, {Hollweg}, {Horbury},
  {Howard}, {Hoxie}, {Jannet}, {Karlsson}, {Kasper}, {Kellogg}, {Kien},
  {Klimchuk}, {Krasnoselskikh}, {Krucker}, {Lynch}, {Maksimovic}, {Malaspina},
  {Marker}, {Martin}, {Martinez-Oliveros}, {McCauley}, {McComas}, {McDonald},
  {Meyer-Vernet}, {Moncuquet}, {Monson}, {Mozer}, {Murphy}, {Odom},
  {Oliverson}, {Olson}, {Parker}, {Pankow}, {Phan}, {Quataert}, {Quinn},
  {Ruplin}, {Salem}, {Seitz}, {Sheppard}, {Siy}, {Stevens}, {Summers}, {Szabo},
  {Timofeeva}, {Vaivads}, {Velli}, {Yehle}, {Werthimer}, \&
  {Wygant}}]{Bale2016}
{Bale}, S.~D., {Goetz}, K., {Harvey}, P.~R., {et~al.} 2016, Space Science Rev.,
  204, 49, \dodoi{10.1007/s11214-016-0244-5}

\bibitem[{{Bandyopadhyay} {et~al.}(2018){Bandyopadhyay}, {Oughton}, {Wan},
  {Matthaeus}, {Chhiber}, \& {Parashar}}]{Bandyopadhyay2018}
{Bandyopadhyay}, R., {Oughton}, S., {Wan}, M., {et~al.} 2018, Physical Review
  X, 8, 041052, \dodoi{10.1103/PhysRevX.8.041052}

\bibitem[{{Bandyopadhyay} {et~al.}(2020){Bandyopadhyay}, {Goldstein}, {Maruca},
  {Matthaeus}, {Parashar}, {Ruffolo}, {Chhiber}, {Usmanov}, {Chasapis},
  {Qudsi}, {Bale}, {Bonnell}, {Dudok de Wit}, {Goetz}, {Harvey}, {MacDowall},
  {Malaspina}, {Pulupa}, {Kasper}, {Korreck}, {Case}, {Stevens}, {Whittlesey},
  {Larson}, {Livi}, {Klein}, {Velli}, \& {Raouafi}}]{Bandyopadhyay2020}
{Bandyopadhyay}, R., {Goldstein}, M.~L., {Maruca}, B.~A., {et~al.} 2020, ApJs,
  246, 48, \dodoi{10.3847/1538-4365/ab5dae}

\bibitem[{{Bavassano} {et~al.}(1998){Bavassano}, {Pietropaolo}, \&
  {Bruno}}]{Bavassano1998}
{Bavassano}, B., {Pietropaolo}, E., \& {Bruno}, R. 1998, JGR, 103, 6521,
  \dodoi{10.1029/97JA03029}

\bibitem[{{Boardsen} {et~al.}(2015){Boardsen}, {Jian}, {Raines}, {Gershman},
  {Zurbuchen}, {Roberts}, \& {Korth}}]{Boardsen2015}
{Boardsen}, S.~A., {Jian}, L.~K., {Raines}, J.~L., {et~al.} 2015, Journal of
  Geophysical Research (Space Physics), 120, 10,207,
  \dodoi{10.1002/2015JA021506}

\bibitem[{{Bowen} {et~al.}(2023){Bowen}, {Chandran}, {Klein}, {Mallet}, {Bale},
  {Squire}, \& {Verniero}}]{Bowen2023}
{Bowen}, T.~A., {Chandran}, B.~D.~G., {Klein}, K.~G., {et~al.} 2023, in 2023
  XXXVth General Assembly and Scientific Symposium of the International Union
  of Radio Science (URSI GASS, 335,
  \dodoi{10.23919/URSIGASS57860.2023.10265538}

\bibitem[{{Bowen} {et~al.}(2020{\natexlab{a}}){Bowen}, {Mallet}, {Huang},
  {Klein}, {Malaspina}, {Stevens}, {Bale}, {Bonnell}, {Case}, {Chandran},
  {Chaston}, {Chen}, {Dudok de Wit}, {Goetz}, {Harvey}, {Howes}, {Kasper},
  {Korreck}, {Larson}, {Livi}, {MacDowall}, {McManus}, {Pulupa}, {Verniero}, \&
  {Whittlesey}}]{Bowen2020a}
{Bowen}, T.~A., {Mallet}, A., {Huang}, J., {et~al.} 2020{\natexlab{a}}, ApJS,
  246, 66, \dodoi{10.3847/1538-4365/ab6c65}

\bibitem[{{Bowen} {et~al.}(2020{\natexlab{b}}){Bowen}, {Bale}, {Bonnell},
  {Larson}, {Mallet}, {McManus}, {Mozer}, {Pulupa}, {Vasko}, {Verniero},
  {Psp/Fields Team}, \& {Psp/Sweap Teams}}]{Bowen2020d}
{Bowen}, T.~A., {Bale}, S.~D., {Bonnell}, J.~W., {et~al.} 2020{\natexlab{b}},
  ApJ, 899, 74, \dodoi{10.3847/1538-4357/ab9f37}

\bibitem[{{Bowen} {et~al.}(2020{\natexlab{c}}){Bowen}, {Mallet}, {Bale},
  {Bonnell}, {Case}, {Chandran}, {Chasapis}, {Chen}, {Duan}, {Dudok de Wit},
  {Goetz}, {Halekas}, {Harvey}, {Kasper}, {Korreck}, {Larson}, {Livi},
  {MacDowall}, {Malaspina}, {McManus}, {Pulupa}, {Stevens}, \&
  {Whittlesey}}]{Bowen2020c}
{Bowen}, T.~A., {Mallet}, A., {Bale}, S.~D., {et~al.} 2020{\natexlab{c}}, PRL,
  125, 025102, \dodoi{10.1103/PhysRevLett.125.025102}

\bibitem[{{Bowen} {et~al.}(2020{\natexlab{d}}){Bowen}, {Bale}, {Bonnell},
  {Dudok de Wit}, {Goetz}, {Goodrich}, {Gruesbeck}, {Harvey}, {Jannet},
  {Koval}, {MacDowall}, {Malaspina}, {Pulupa}, {Revillet}, {Sheppard}, \&
  {Szabo}}]{Bowen2020b}
{Bowen}, T.~A., {Bale}, S.~D., {Bonnell}, J.~W., {et~al.} 2020{\natexlab{d}},
  Journal of Geophysical Research (Space Physics), 125, e27813,
  \dodoi{10.1029/2020JA027813}

\bibitem[{{Bowen} {et~al.}(2022){Bowen}, {Chandran}, {Squire}, {Bale}, {Duan},
  {Klein}, {Larson}, {Mallet}, {McManus}, {Meyrand}, {Verniero}, \&
  {Woodham}}]{Bowen2022}
{Bowen}, T.~A., {Chandran}, B. D.~G., {Squire}, J., {et~al.} 2022, PRL, 129,
  165101, \dodoi{10.1103/PhysRevLett.129.165101}

\bibitem[{Broomhead \& Lowe(1988)}]{broomhead1988}
Broomhead, D., \& Lowe, D. 1988, Complex Systems, 2, 321

\bibitem[{{Bruno} {et~al.}(2014){Bruno}, {Trenchi}, \& {Telloni}}]{Bruno2014}
{Bruno}, R., {Trenchi}, L., \& {Telloni}, D. 2014, ApJL, 793, L15,
  \dodoi{10.1088/2041-8205/793/1/L15}

\bibitem[{{Chandran} {et~al.}(2010){Chandran}, {Pongkitiwanichakul},
  {Isenberg}, {Lee}, {Markovskii}, {Hollweg}, \& {Vasquez}}]{Chandran2010b}
{Chandran}, B. D.~G., {Pongkitiwanichakul}, P., {Isenberg}, P.~A., {et~al.}
  2010, ApJ, 722, 710, \dodoi{10.1088/0004-637X/722/1/710}

\bibitem[{{Chen} {et~al.}(2010){Chen}, {Horbury}, {Schekochihin}, {Wicks},
  {Alexandrova}, \& {Mitchell}}]{Chen2010a}
{Chen}, C.~H.~K., {Horbury}, T.~S., {Schekochihin}, A.~A., {et~al.} 2010,
  Physical Review Letters, 104, 255002, \dodoi{10.1103/PhysRevLett.104.255002}

\bibitem[{{Chew} {et~al.}(1956){Chew}, {Goldberger}, \& {Low}}]{Chew1956}
{Chew}, G.~F., {Goldberger}, M.~L., \& {Low}, F.~E. 1956, Proceedings of the
  Royal Society of London Series A, 236, 112, \dodoi{10.1098/rspa.1956.0116}

\bibitem[{{Cranmer}(2000)}]{Cranmer2000}
{Cranmer}, S.~R. 2000, The Astrophysical Journal, 532, 1197,
  \dodoi{10.1086/308620}

\bibitem[{{Cranmer}(2014)}]{Cranmer2014}
---. 2014, ApJS, 213, 16, \dodoi{10.1088/0067-0049/213/1/16}

\bibitem[{{Davis} {et~al.}(2023){Davis}, {Chandran}, {Bowen}, {Badman}, {Dudok
  de Wit}, {Chen}, {Bale}, {Huang}, {Sioulas}, \& {Velli}}]{Davis2023}
{Davis}, N., {Chandran}, B.~D.~G., {Bowen}, T.~A., {et~al.} 2023, arXiv
  e-prints, arXiv:2303.01663, \dodoi{10.48550/arXiv.2303.01663}

\bibitem[{{Denskat} {et~al.}(1983){Denskat}, {Beinroth}, \&
  {Neubauer}}]{Denskat1983}
{Denskat}, K.~U., {Beinroth}, H.~J., \& {Neubauer}, F.~M. 1983, Journal of
  Geophysics Zeitschrift Geophysik, 54, 60

\bibitem[{{Duan} {et~al.}(2021){Duan}, {He}, {Bowen}, {Woodham}, {Wang},
  {Chen}, {Mallet}, \& {Bale}}]{Duan2021}
{Duan}, D., {He}, J., {Bowen}, T.~A., {et~al.} 2021, ApJL, 915, L8,
  \dodoi{10.3847/2041-8213/ac07ac}

\bibitem[{{Duan} {et~al.}(2020){Duan}, {Bowen}, {Chen}, {Mallet}, {He}, {Bale},
  {Vech}, {Kasper}, {Pulupa}, {Bonnell}, {Case}, {de Wit}, {Goetz}, {Harvey},
  {Korreck}, {Larson}, {Livi}, {MacDowall}, {Malaspina}, {Stevens}, \&
  {Whittlesey}}]{Duan2020}
{Duan}, D., {Bowen}, T.~A., {Chen}, C. H.~K., {et~al.} 2020, ApJS, 246, 55,
  \dodoi{10.3847/1538-4365/ab672d}

\bibitem[{{Dudok de Wit} {et~al.}(2013){Dudok de Wit}, {Alexandrova}, {Furno},
  {Sorriso-Valvo}, \& {Zimbardo}}]{DudokdeWit2013}
{Dudok de Wit}, T., {Alexandrova}, O., {Furno}, I., {Sorriso-Valvo}, L., \&
  {Zimbardo}, G. 2013, \ssr, 178, 665, \dodoi{10.1007/s11214-013-9974-9}

\bibitem[{{Dudok de Wit} {et~al.}(2022){Dudok de Wit}, {Krasnoselskikh},
  {Agapitov}, {Froment}, {Larosa}, {Bale}, {Bowen}, {Goetz}, {Harvey},
  {Jannet}, {Kretzschmar}, {MacDowall}, {Malaspina}, {Martin}, {Page},
  {Pulupa}, \& {Revillet}}]{DudokdeWit2022}
{Dudok de Wit}, T., {Krasnoselskikh}, V.~V., {Agapitov}, O., {et~al.} 2022,
  Journal of Geophysical Research (Space Physics), 127, e30018,
  \dodoi{10.1029/2021JA030018}

\bibitem[{{Dum} {et~al.}(1980){Dum}, {Marsch}, \& {Pilipp}}]{Dum1980}
{Dum}, C.~T., {Marsch}, E., \& {Pilipp}, W. 1980, Journal of Plasma Physics,
  23, 91, \dodoi{10.1017/S0022377800022170}

\bibitem[{{Farge}(1992)}]{Farge1992}
{Farge}, M. 1992, Annual Review of Fluid Mechanics, 24, 395,
  \dodoi{10.1146/annurev.fl.24.010192.002143}

\bibitem[{Fox {et~al.}(2016)Fox, Velli, Bale, Decker, Driesman, Howard, Kasper,
  Kinnison, Kusterer, Lario, Lockwood, McComas, Raouafi, \& Szabo}]{Fox2016}
Fox, N.~J., Velli, M.~C., Bale, S.~D., {et~al.} 2016, Space Science Reviews,
  204, 7, \dodoi{10.1007/s11214-015-0211-6}

\bibitem[{{Gary}(1993)}]{Gary1993}
{Gary}, S.~P. 1993, {Theory of Space Plasma Microinstabilities}

\bibitem[{{Goldstein} {et~al.}(1994){Goldstein}, {Roberts}, \&
  {Fitch}}]{Goldstein1994}
{Goldstein}, M.~L., {Roberts}, D.~A., \& {Fitch}, C.~A. 1994, Journal of
  Geophysical Research, 99, 11519, \dodoi{10.1029/94JA00789}

\bibitem[{{He} {et~al.}(2011){He}, {Marsch}, {Tu}, {Yao}, \& {Tian}}]{He2011}
{He}, J., {Marsch}, E., {Tu}, C., {Yao}, S., \& {Tian}, H. 2011, The
  Astrophysical Journal, 731, 85, \dodoi{10.1088/0004-637X/731/2/85}

\bibitem[{{He} {et~al.}(2015){He}, {Wang}, {Tu}, {Marsch}, \& {Zong}}]{He2015}
{He}, J., {Wang}, L., {Tu}, C., {Marsch}, E., \& {Zong}, Q. 2015, ApJ, 800,
  L31, \dodoi{10.1088/2041-8205/800/2/L31}

\bibitem[{{Hellinger} {et~al.}(2013){Hellinger}, {Tr{\'a}Vn{\'\i}{\v{c}}ek},
  {{\v{S}}tver{\'a}k}, {Matteini}, \& {Velli}}]{Hellinger2013}
{Hellinger}, P., {Tr{\'a}Vn{\'\i}{\v{c}}ek}, P.~M., {{\v{S}}tver{\'a}k},
  {\v{S}}., {Matteini}, L., \& {Velli}, M. 2013, Journal of Geophysical
  Research (Space Physics), 118, 1351, \dodoi{10.1002/jgra.50107}

\bibitem[{{Heuer} \& {Marsch}(2007)}]{Heuer2007}
{Heuer}, M., \& {Marsch}, E. 2007, Journal of Geophysical Research (Space
  Physics), 112, A03102, \dodoi{10.1029/2006JA011979}

\bibitem[{{Hollweg} \& {Isenberg}(2002)}]{HollwegIsenberg2002}
{Hollweg}, J.~V., \& {Isenberg}, P.~A. 2002, Journal of Geophysical Research
  (Space Physics), 107, 1147, \dodoi{10.1029/2001JA000270}

\bibitem[{{Hollweg} \& {Johnson}(1988)}]{HollwegJohnson1988}
{Hollweg}, J.~V., \& {Johnson}, W. 1988, JGR, 93, 9547,
  \dodoi{10.1029/JA093iA09p09547}

\bibitem[{{Hollweg} \& {Markovskii}(2002)}]{Hollweg2002}
{Hollweg}, J.~V., \& {Markovskii}, S.~A. 2002, Journal of Geophysical Research
  (Space Physics), 107, 1080, \dodoi{10.1029/2001JA000205}

\bibitem[{{Horbury} {et~al.}(2008){Horbury}, {Forman}, \&
  {Oughton}}]{Horbury2008}
{Horbury}, T.~S., {Forman}, M., \& {Oughton}, S. 2008, PRL, 101, 175005,
  \dodoi{10.1103/PhysRevLett.101.175005}

\bibitem[{{Hossain} {et~al.}(1995){Hossain}, {Gray}, {Pontius}, {Matthaeus}, \&
  {Oughton}}]{Hossain1995}
{Hossain}, M., {Gray}, P.~C., {Pontius}, Duane~H., J., {Matthaeus}, W.~H., \&
  {Oughton}, S. 1995, Physics of Fluids, 7, 2886, \dodoi{10.1063/1.868665}

\bibitem[{{Howes} \& {Quataert}(2010)}]{HowesQuataert2010}
{Howes}, G.~G., \& {Quataert}, E. 2010, The Astrophysical Journal Letters, 709,
  L49, \dodoi{10.1088/2041-8205/709/1/L49}

\bibitem[{{Huang} {et~al.}(2020){Huang}, {Zhang}, {Sahraoui}, {He}, {Yuan},
  {Andr{\'e}s}, {Hadid}, {Deng}, {Jiang}, {Yu}, {Xiong}, {Wei}, {Xu}, {Bale},
  \& {Kasper}}]{Huang2020}
{Huang}, S.~Y., {Zhang}, J., {Sahraoui}, F., {et~al.} 2020, ApJL, 897, L3,
  \dodoi{10.3847/2041-8213/ab9abb}

\bibitem[{{Isenberg} \& {Lee}(1996)}]{IsenbergLee1996}
{Isenberg}, P.~A., \& {Lee}, M.~A. 1996, JGR, 101, 11055,
  \dodoi{10.1029/96JA00293}

\bibitem[{{Jannet} {et~al.}(2021){Jannet}, {Dudok de Wit}, {Krasnoselskikh},
  {Kretzschmar}, {Fergeau}, {Bergerard-Timofeeva}, {Agrapart}, {Brochot},
  {Chalumeau}, {Martin}, {Revillet}, {Bale}, {Maksimovic}, {Bowen},
  {Brysbaert}, {Goetz}, {Guilhem}, {Harvey}, {Leray}, \&
  {Lorf{\`e}vre}}]{Jannet2021}
{Jannet}, G., {Dudok de Wit}, T., {Krasnoselskikh}, V., {et~al.} 2021, Journal
  of Geophysical Research (Space Physics), 126, e28543,
  \dodoi{10.1029/2020JA028543}

\bibitem[{{Jian} {et~al.}(2014){Jian}, {Wei}, {Russell}, {Luhmann}, {Klecker},
  {Omidi}, {Isenberg}, {Goldstein}, {Figueroa-Vi{\~n}as}, \&
  {Blanco-Cano}}]{Jian2014}
{Jian}, L.~K., {Wei}, H.~Y., {Russell}, C.~T., {et~al.} 2014, \apj, 786, 123,
  \dodoi{10.1088/0004-637X/786/2/123}

\bibitem[{{Karpman}(1974)}]{Karpman1974}
{Karpman}, V.~I. 1974, SSR, 16, 361, \dodoi{10.1007/BF00171564}

\bibitem[{{Kennel} \& {Engelmann}(1966)}]{KennelEngelmann1966}
{Kennel}, C.~F., \& {Engelmann}, F. 1966, Physics of Fluids, 9, 2377,
  \dodoi{10.1063/1.1761629}

\bibitem[{{Kiyani} {et~al.}(2009){Kiyani}, {Chapman}, {Khotyaintsev}, {Dunlop},
  \& {Sahraoui}}]{Kiyani2009}
{Kiyani}, K.~H., {Chapman}, S.~C., {Khotyaintsev}, Y.~V., {Dunlop}, M.~W., \&
  {Sahraoui}, F. 2009, Physical Review Letters, 103, 075006,
  \dodoi{10.1103/PhysRevLett.103.075006}

\bibitem[{{Klein} {et~al.}(2018){Klein}, {Alterman}, {Stevens}, {Vech}, \&
  {Kasper}}]{Klein2018}
{Klein}, K.~G., {Alterman}, B.~L., {Stevens}, M.~L., {Vech}, D., \& {Kasper},
  J.~C. 2018, PRL, 120, 205102, \dodoi{10.1103/PhysRevLett.120.205102}

\bibitem[{{Klein} {et~al.}(2021){Klein}, {Verniero}, {Alterman}, {Bale},
  {Case}, {Kasper}, {Korreck}, {Larson}, {Lichko}, {Livi}, {McManus},
  {Martinovi{\'c}}, {Rahmati}, {Stevens}, \& {Whittlesey}}]{Klein2021}
{Klein}, K.~G., {Verniero}, J.~L., {Alterman}, B., {et~al.} 2021, ApJ, 909, 7,
  \dodoi{10.3847/1538-4357/abd7a0}

\bibitem[{{Leamon} {et~al.}(1998){Leamon}, {Smith}, {Ness}, {Matthaeus}, \&
  {Wong}}]{Leamon1998a}
{Leamon}, R.~J., {Smith}, C.~W., {Ness}, N.~F., {Matthaeus}, W.~H., \& {Wong},
  H.~K. 1998, Journal of Geophysical Research, 103, 4775,
  \dodoi{10.1029/97JA03394}

\bibitem[{{Lion} {et~al.}(2016){Lion}, {Alexandrova}, \&
  {Zaslavsky}}]{Lion2016}
{Lion}, S., {Alexandrova}, O., \& {Zaslavsky}, A. 2016, The Astrophysical
  Journal, 824, 47, \dodoi{10.3847/0004-637X/824/1/47}

\bibitem[{{Liu} {et~al.}(2023){Liu}, {Zhao}, {Wang}, {Dong}, {Kasper}, {Bale},
  {Shi}, \& {Wu}}]{Liu2023}
{Liu}, W., {Zhao}, J., {Wang}, T., {et~al.} 2023, \apj, 951, 69,
  \dodoi{10.3847/1538-4357/acd53b}

\bibitem[{{Livi} {et~al.}(2022){Livi}, {Larson}, {Kasper}, {Abiad}, {Case},
  {Klein}, {Curtis}, {Dalton}, {Stevens}, {Korreck}, {Ho}, {Robinson}, {Tiu},
  {Whittlesey}, {Verniero}, {Halekas}, {McFadden}, {Marckwordt}, {Slagle},
  {Abatcha}, {Rahmati}, \& {McManus}}]{Livi2022}
{Livi}, R., {Larson}, D.~E., {Kasper}, J.~C., {et~al.} 2022, ApJ, 938, 138,
  \dodoi{10.3847/1538-4357/ac93f5}

\bibitem[{{Lotz} {et~al.}(2023){Lotz}, {Nel}, {Wicks}, {Roberts},
  {Engelbrecht}, {Strauss}, {Botha}, {Kontar}, {Pit{\v{n}}a}, \&
  {Bale}}]{Lotz2023}
{Lotz}, S., {Nel}, A.~E., {Wicks}, R.~T., {et~al.} 2023, ApJ, 942, 93,
  \dodoi{10.3847/1538-4357/aca903}

\bibitem[{{Luo} {et~al.}(2022){Luo}, {Zhu}, {He}, {Cui}, {Lai}, {Verscharen},
  \& {Duan}}]{Luo2022}
{Luo}, Q., {Zhu}, X., {He}, J., {et~al.} 2022, \apj, 928, 36,
  \dodoi{10.3847/1538-4357/ac52a9}

\bibitem[{{Marsch}(2006)}]{Marsch2006}
{Marsch}, E. 2006, Living Reviews in Solar Physics, 3, 1,
  \dodoi{10.12942/lrsp-2006-1}

\bibitem[{{Marsch} \& {Goldstein}(1983)}]{MarschGoldstein1983}
{Marsch}, E., \& {Goldstein}, H. 1983, \jgr, 88, 9933,
  \dodoi{10.1029/JA088iA12p09933}

\bibitem[{{Marsch} {et~al.}(1982){Marsch}, {Schwenn}, {Rosenbauer},
  {Muehlhaeuser}, {Pilipp}, \& {Neubauer}}]{Marsch1982}
{Marsch}, E., {Schwenn}, R., {Rosenbauer}, H., {et~al.} 1982, \jgr, 87, 52,
  \dodoi{10.1029/JA087iA01p00052}

\bibitem[{{Marsch} \& {Tu}(2001{\natexlab{a}})}]{MarschTu2001b}
{Marsch}, E., \& {Tu}, C.~Y. 2001{\natexlab{a}}, JGR, 106, 8357,
  \dodoi{10.1029/2000JA000414}

\bibitem[{{Marsch} \& {Tu}(2001{\natexlab{b}})}]{MarschTu2001a}
---. 2001{\natexlab{b}}, JGR, 106, 227, \dodoi{10.1029/2000JA000042}

\bibitem[{{McManus} {et~al.}(2020){McManus}, {Bowen}, {Mallet}, {Chen},
  {Chandran}, {Bale}, {Larson}, {Dudok de Wit}, {Kasper}, {Stevens},
  {Whittlesey}, {Livi}, {Korreck}, {Goetz}, {Harvey}, {Pulupa}, {MacDowall},
  {Malaspina}, {Case}, \& {Bonnell}}]{McManus2020}
{McManus}, M.~D., {Bowen}, T.~A., {Mallet}, A., {et~al.} 2020, ApJS, 246, 67,
  \dodoi{10.3847/1538-4365/ab6dce}

\bibitem[{{McManus} {et~al.}(2023){McManus}, {Klein}, {Larson}, {Bale},
  {Bowen}, {Huang}, {Livi}, {Rahmati}, {Romeo}, {Verniero}, \&
  {Whittlesey}}]{McManus2023}
{McManus}, M.~D., {Klein}, K.~G., {Larson}, D., {et~al.} 2023, arXiv e-prints,
  arXiv:2310.14136, \dodoi{10.48550/arXiv.2310.14136}

\bibitem[{{Meyrand} {et~al.}(2021){Meyrand}, {Squire}, {Schekochihin}, \&
  {Dorland}}]{Meyrand2021}
{Meyrand}, R., {Squire}, J., {Schekochihin}, A.~A., \& {Dorland}, W. 2021,
  Journal of Plasma Physics, 87, 535870301, \dodoi{10.1017/S0022377821000489}

\bibitem[{{Ofman} {et~al.}(2022){Ofman}, {Boardsen}, {Jian}, {Verniero}, \&
  {Larson}}]{Ofman2022}
{Ofman}, L., {Boardsen}, S.~A., {Jian}, L.~K., {Verniero}, J.~L., \& {Larson},
  D. 2022, \apj, 926, 185, \dodoi{10.3847/1538-4357/ac402c}

\bibitem[{{Parker}(1958)}]{Parker1958}
{Parker}, E.~N. 1958, ApJ, 128, 664, \dodoi{10.1086/146579}

\bibitem[{{Podesta} \& {Gary}(2011)}]{Podesta2011}
{Podesta}, J.~J., \& {Gary}, S.~P. 2011, The Astrophysical Journal, 734, 15,
  \dodoi{10.1088/0004-637X/734/1/15}

\bibitem[{{Richardson} {et~al.}(1995){Richardson}, {Paularena}, {Lazarus}, \&
  {Belcher}}]{Richardson1995}
{Richardson}, J.~D., {Paularena}, K.~I., {Lazarus}, A.~J., \& {Belcher}, J.~W.
  1995, Geophys. Research Letters, 22, 325, \dodoi{10.1029/94GL03273}

\bibitem[{{Roberts} {et~al.}(1987){Roberts}, {Klein}, {Goldstein}, \&
  {Matthaeus}}]{Roberts1987a}
{Roberts}, D.~A., {Klein}, L.~W., {Goldstein}, M.~L., \& {Matthaeus}, W.~H.
  1987, JGR, 92, 11021, \dodoi{10.1029/JA092iA10p11021}

\bibitem[{{Sahraoui} {et~al.}(2009){Sahraoui}, {Goldstein}, {Robert}, \&
  {Khotyaintsev}}]{Sahraoui2009}
{Sahraoui}, F., {Goldstein}, M.~L., {Robert}, P., \& {Khotyaintsev}, Y.~V.
  2009, Physical Review Letters, 102, 231102,
  \dodoi{10.1103/PhysRevLett.102.231102}

\bibitem[{{Shankarappa} {et~al.}(2023){Shankarappa}, {Klein}, \&
  {Martinovi{\'c}}}]{Shankarappa2023}
{Shankarappa}, N., {Klein}, K.~G., \& {Martinovi{\'c}}, M.~M. 2023, ApJ, 946,
  85, \dodoi{10.3847/1538-4357/acb542}

\bibitem[{{Shebalin} {et~al.}(1983){Shebalin}, {Matthaeus}, \&
  {Montgomery}}]{Shebalin1983}
{Shebalin}, J.~V., {Matthaeus}, W.~H., \& {Montgomery}, D. 1983, Journal of
  Plasma Physics, 29, 525, \dodoi{10.1017/S0022377800000933}

\bibitem[{{Shklyar} \& {Matsumoto}(2009)}]{Shklyar}
{Shklyar}, D., \& {Matsumoto}, H. 2009, Surveys in Geophysics, 30, 55,
  \dodoi{10.1007/s10712-009-9061-7}

\bibitem[{{Smith} {et~al.}(2006){Smith}, {Hamilton}, {Vasquez}, \&
  {Leamon}}]{Smith2006}
{Smith}, C.~W., {Hamilton}, K., {Vasquez}, B.~J., \& {Leamon}, R.~J. 2006, The
  Astrophysical Journal Letters, 645, L85, \dodoi{10.1086/506151}

\bibitem[{{Smith} {et~al.}(2009){Smith}, {Stawarz}, {Vasquez}, {Forman}, \&
  {MacBride}}]{Smith2009}
{Smith}, C.~W., {Stawarz}, J.~E., {Vasquez}, B.~J., {Forman}, M.~A., \&
  {MacBride}, B.~T. 2009, PRL, 103, 201101,
  \dodoi{10.1103/PhysRevLett.103.201101}

\bibitem[{{Smith} {et~al.}(2012){Smith}, {Vasquez}, \& {Hollweg}}]{Smith2012}
{Smith}, C.~W., {Vasquez}, B.~J., \& {Hollweg}, J.~V. 2012, ApJ, 745, 8,
  \dodoi{10.1088/0004-637X/745/1/8}

\bibitem[{{Smith} {et~al.}(1990){Smith}, {Matthaeus}, \& {Ness}}]{Smith1990}
{Smith}, W.~C., {Matthaeus}, H.~W., \& {Ness}, F.~N. 1990, in International
  Cosmic Ray Conference, Vol.~5, International Cosmic Ray Conference, 280

\bibitem[{{Squire} {et~al.}(2022){Squire}, {Meyrand}, {Kunz}, {Arzamasskiy},
  {Schekochihin}, \& {Quataert}}]{Squire2022}
{Squire}, J., {Meyrand}, R., {Kunz}, M.~W., {et~al.} 2022, Nature Astronomy, 6,
  715, \dodoi{10.1038/s41550-022-01624-z}

\bibitem[{{Stix}(1992)}]{Stix1992}
{Stix}, T.~H. 1992, {Waves in plasmas}

\bibitem[{{Telloni} {et~al.}(2019){Telloni}, {Carbone}, {Bruno}, {Zank},
  {Sorriso-Valvo}, \& {Mancuso}}]{Telloni2019}
{Telloni}, D., {Carbone}, F., {Bruno}, R., {et~al.} 2019, ApJL, 885, L5,
  \dodoi{10.3847/2041-8213/ab4c44}

\bibitem[{{Tu} \& {Marsch}(1995)}]{TuMarsch1995}
{Tu}, C.~Y., \& {Marsch}, E. 1995, Space Sci Rev, 73, 1,
  \dodoi{10.1007/BF00748891}

\bibitem[{{Tu} \& {Marsch}(1997)}]{TuMarsch1997}
---. 1997, Sol Phys, 171, 363, \dodoi{10.1023/A:1004968327196}

\bibitem[{{Tu} \& {Marsch}(2001)}]{TuMarsch2001}
---. 2001, JGR, 106, 8233, \dodoi{10.1029/2000JA000024}

\bibitem[{{Usmanov} {et~al.}(2014){Usmanov}, {Goldstein}, \&
  {Matthaeus}}]{Usmanov2014}
{Usmanov}, A.~V., {Goldstein}, M.~L., \& {Matthaeus}, W.~H. 2014, \apj, 788,
  43, \dodoi{10.1088/0004-637X/788/1/43}

\bibitem[{{Vech} {et~al.}(2018){Vech}, {Mallet}, {Klein}, \&
  {Kasper}}]{Vech2018}
{Vech}, D., {Mallet}, A., {Klein}, K.~G., \& {Kasper}, J.~C. 2018, The
  Astrophysical Journal Letters, 855, L27, \dodoi{10.3847/2041-8213/aab351}

\bibitem[{{Vech} {et~al.}(2020){Vech}, {Kasper}, {Klein}, {Huang}, {Stevens},
  {Chen}, {Case}, {Korreck}, {Bale}, {Bowen}, {Whittlesey}, {Livi}, {Larson},
  {Malaspina}, {Pulupa}, {Bonnell}, {Harvey}, {Goetz}, {Dudok de Wit}, \&
  {MacDowall}}]{Vech2020}
{Vech}, D., {Kasper}, J.~C., {Klein}, K.~G., {et~al.} 2020, ApJS, 246, 52,
  \dodoi{10.3847/1538-4365/ab60a2}

\bibitem[{{Verniero} {et~al.}(2020){Verniero}, {Larson}, {Livi}, {Rahmati},
  {McManus}, {Pyakurel}, {Klein}, {Bowen}, {Bonnell}, {Alterman}, {Whittlesey},
  {Malaspina}, {Bale}, {Kasper}, {Case}, {Goetz}, {Harvey}, {Korreck},
  {MacDowall}, {Pulupa}, {Stevens}, \& {de Wit}}]{Verniero2020}
{Verniero}, J.~L., {Larson}, D.~E., {Livi}, R., {et~al.} 2020, ApJS, 248, 5,
  \dodoi{10.3847/1538-4365/ab86af}

\bibitem[{{Verniero} {et~al.}(2022){Verniero}, {Chandran}, {Larson}, {Paulson},
  {Alterman}, {Badman}, {Bale}, {Bonnell}, {Bowen}, {de Wit}, {Kasper},
  {Klein}, {Lichko}, {Livi}, {McManus}, {Rahmati}, {Verscharen}, {Walters}, \&
  {Whittlesey}}]{Verniero2021}
{Verniero}, J.~L., {Chandran}, B.~D.~G., {Larson}, D.~E., {et~al.} 2022, ApJ,
  924, 112, \dodoi{10.3847/1538-4357/ac36d5}

\bibitem[{{Verscharen} {et~al.}(2017){Verscharen}, {Chen}, \&
  {Wicks}}]{Verscharen2017}
{Verscharen}, D., {Chen}, C.~H.~K., \& {Wicks}, R.~T. 2017, ApJ, 840, 106,
  \dodoi{10.3847/1538-4357/aa6a56}

\bibitem[{{Vi{\~n}as} \& {Gurgiolo}(2009)}]{Vinas2009}
{Vi{\~n}as}, A.~F., \& {Gurgiolo}, C. 2009, Journal of Geophysical Research
  (Space Physics), 114, A01105, \dodoi{10.1029/2008JA013633}

\bibitem[{{Walters} {et~al.}(2023){Walters}, {Klein}, {Lichko}, {Stevens},
  {Verscharen}, \& {Chandran}}]{Walters2023}
{Walters}, J., {Klein}, K.~G., {Lichko}, E., {et~al.} 2023, \apj, 955, 97,
  \dodoi{10.3847/1538-4357/acf1fa}

\bibitem[{{Wan} {et~al.}(2012){Wan}, {Matthaeus}, {Karimabadi}, {Roytershteyn},
  {Shay}, {Wu}, {Daughton}, {Loring}, \& {Chapman}}]{Wan2012}
{Wan}, M., {Matthaeus}, W.~H., {Karimabadi}, H., {et~al.} 2012, PRL, 109,
  195001, \dodoi{10.1103/PhysRevLett.109.195001}

\bibitem[{{Wicks} {et~al.}(2016){Wicks}, {Alexander}, {Stevens}, {Wilson},
  {Moya}, {Vi{\~n}as}, {Jian}, {Roberts}, {O'Modhrain}, {Gilbert}, \&
  {Zurbuchen}}]{Wicks2016}
{Wicks}, R.~T., {Alexander}, R.~L., {Stevens}, M., {et~al.} 2016, ApJ, 819, 6,
  \dodoi{10.3847/0004-637X/819/1/6}

\bibitem[{{Woodham} {et~al.}(2018){Woodham}, {Wicks}, {Verscharen}, \&
  {Owen}}]{Woodham2018}
{Woodham}, L.~D., {Wicks}, R.~T., {Verscharen}, D., \& {Owen}, C.~J. 2018, The
  Astrophysical Journal, 856, 49, \dodoi{10.3847/1538-4357/aab03d}

\bibitem[{{Woodham} {et~al.}(2019){Woodham}, {Wicks}, {Verscharen}, {Owen},
  {Maruca}, \& {Alterman}}]{Woodham2019}
{Woodham}, L.~D., {Wicks}, R.~T., {Verscharen}, D., {et~al.} 2019, ApJL, 884,
  L53, \dodoi{10.3847/2041-8213/ab4adc}

\bibitem[{{Wu} {et~al.}(2022){Wu}, {Tu}, {He}, {Wang}, \& {Yang}}]{Wu2022}
{Wu}, H., {Tu}, C., {He}, J., {Wang}, X., \& {Yang}, L. 2022, ApJ, 926, 116,
  \dodoi{10.3847/1538-4357/ac4413}

\bibitem[{{W{\"u}est} {et~al.}(2007){W{\"u}est}, {Evans}, \& {von
  Steiger}}]{Wuest2007}
{W{\"u}est}, M., {Evans}, D.~S., \& {von Steiger}, R. 2007, {Calibration of
  Particle Instruments in Space Physics}

\bibitem[{{Zhao} {et~al.}(2021){Zhao}, {Lin}, {Wang}, {Feng}, {Wu}, {Li},
  {Zhao}, \& {Liu}}]{Zhao2021}
{Zhao}, G.~Q., {Lin}, Y., {Wang}, X.~Y., {et~al.} 2021, ApJ, 906, 123,
  \dodoi{10.3847/1538-4357/abca3b}

\end{thebibliography}

\appendix
\section{Numerical Determination of Quasilinear Heating Rate}\label{AppendA}
Determination of $Q_{\rm{ICW}}$ and $Q(v_\perp,v_\parallel)$ is obtained through integrating  Eq \ref{eq:H} by parts to obtain 

\begin{align} \label{eq:QQ}
Q_{\rm ICW} =-\pi^2\Omega_p^2m_p\int_{-\infty}^{\infty} dv_\parallel \int_0^\infty  dv_\perp \Bigg[\int_0^\infty dk_\parallel  v_\perp^2 \delta(\omega_k -k_\parallel v_\parallel-\Omega_p)\frac{\omega_k^2}{k_\parallel^2} I(k_\parallel) \hat{G}_k \bar{g}(v_\perp,v_\parallel)\Bigg]
\end{align}

The $\delta-$function in Eq \ref{eq:H}\& \ref{eq:QQ} corresponds to the ICW resonance condition, and defines a single parallel wave number that resonates with particles at each parallel velocity. Change of variables yields a $\delta$-function for resonant wave number $k_{res}(v_\parallel)$ as \begin{equation}
    \delta(\omega_k -k_\parallel v_\parallel-\Omega_p)=\frac{\delta(k_\parallel-k_{res})}{|v_g(k_{res})-v_\parallel|},\end{equation}
    where $v_g(k_{res})=\partial\omega_k/\partial k|_{k_{res}}$ is the group speed evaluated at each $k_{res}$. For positive $k_\parallel$, taken as outward going ICWs, there is no resonance unless $v_\parallel <0$, limiting our concern only to negative thermal speeds. We use the $\delta-$function to substitute $k_{res}(v_\parallel)$ for $k_\parallel$, thereby evaluating the integral over wavenumber, Eq \ref{eq:QQ} becomes 

\begin{align}
Q_{\rm ICW} =-\pi^2\Omega_p^2m_p\int_{-\infty}^{0} dv_\parallel \int_0^\infty  dv_\perp \Bigg[\frac{v_\perp^2}{|v_g(k_{res})-v_\parallel|}\frac{\omega_{k_{res}}^2}{k_{res}^2} I(k_{res}) \hat{G}_k \bar{g}(v_\perp,v_\parallel)\Bigg],\label{eq:QQ2} \end{align}
where the spectrum $I(k_\parallel)$ is interpolated from $k_\parallel$ to $k_{res}$, constructing $I(k_{res})$. Eq \ref{eq:QQ2} is entirely a function of $v_\parallel$ and $v_\perp$ and can be calculated via a numerical integration over velocity space. We further identify the differential heating rate \begin{align} 
Q(v_\perp,v_\parallel)=-\pi^2\Omega_p^2m_p\frac{v_\perp^2}{|v_g(k_{res})-v_\parallel|}\frac{\omega_{k_{res}}^2}{k_{res}^2} I(k_{res}) \hat{G}_k \bar{g}(v_\perp,v_\parallel).
\end{align}
To perform the numerical integral in Eq \ref{eq:QQ2}, we evaluate $Q(v_\perp,v_\parallel)$ on a 1 km/s $\times$ 1 km/s grid, which was found to be sufficient for convergence of Eq \ref{eq:QQ2}. The range of integration is set to -1000 \textrm{km/s} $<v_\parallel<$ 0 \textrm{km/s} and 0 \textrm{km/s} $<v_\perp<$  1000 \textrm{km/s}. We have removed contributions to the spectra with $k_\parallel d_i<0.3$ and $k_\parallel d_i>5$, which for $\beta_p=0.1$ corresponds to an approximate range of thermal speeds between 0.02 and 7.5 ${v_{th}}_\parallel$. Figure \ref{fig5}(a\&b) shows that the parallel thermal speeds hovers near 50 km/s over this entire interval. Thus, the 1000km/s limits are sufficiently large with respect to the thermal and Alfv\'{e}n speeds for this stream that there are no significant contributions in portion of phase space out of the bounds of integration. The heating rate as a function of parallel resonant velocity $Q(v_\parallel)$ is computed by integrating $Q(v_\perp,v_\parallel)$ over perpendicular velocities.

These methods have previously been used to obtain heating rates $Q_{\rm ICW}$ of order $\sim 10\%$ of the turbulent cascade rate \citep{Bowen2022} though this previously studied interval contained significantly less amounts of left-hand polarized waves than this stream. Our results shown in Fig \ref{fig5} suggest that a great amount of the turbulent energy likely enters the plasma via cyclotron resonant heating. 

\section{Application of Non-Parametric Representations}
\label{AppendB}
While in situ observations of the proton velocity distribution have long been recognized to have a mostly Gaussian shape \citep{MarschGoldstein1983} and drifting biMaxwellian approximation has regularly been used to approximate ion velocity distributions \citep{Marsch1982,MarschTu2001a,TuMarsch2001,Heuer2007,Klein2021,Verniero2021,McManus2023}; however, it is clear that deviations from nonthermal structure in the velocity distribution can affect wave-particle resonant processes \citep{Dum1980,Bowen2022, Walters2023}. Furthermore, it is well known that the equilibrium kinetic contours for dispersive ICW resonance do not coincide with Maxwellian curvature \citep{IsenbergLee1996}, such that relaxation via ICW resonant processing cannot be entirely captured via bi-Maxwellian approximations \citep{Heuer2007}. In this sense it is important to test the heating rates computed from the drifting biMaxwellian model in Equation \ref{eq:bimax} against heating rates computed from non-parametric models of the velocity distribution. 

\subsection{Hermite Polynomial Interpolation}
\cite{Bowen2022} previously used linear least-square fits of orthogonal Hermite-polynomials to estimate non-Maxwellian features in the velocity distribution; following this previously developed method, we perform a linear least square fit to the average proton velocity distribution in each 128s interval using the Hermite polynomials $H_n$, and Hermite functions  $\phi_m$
\begin{align}
\bar{g}^H(v_\perp, v_\parallel)= \sum_{m,n} A_{mn}\phi^m(v_\perp/w_{\perp{th}})\phi^n(v_\parallel/w_{\parallel{th}})\label{eq:hmte}\\
H_n(v) = (-1)^n e^{v^2}\frac{d^n}{dx^n}e^{-v^2}\\
\phi^m=\frac{H^m(v)}{\sqrt{2^m \pi^{1/2} m!}}e^{-v^2}.
\end{align}

 We fit an orthogonal set of Hermite functions to each velocity distribution through constructing the matrix $\mathbf{A}$, which consists of the transform coefficients $A_{mn}$ corresponding to each two-dimensional Hermite polynomial combination. Through linear least square fitting of $\Phi$, with elements $\phi^m\phi^n$, to the observed velocity distribution, written as matrix $\mathbf{g}$ such that $\mathbf{g}={\mathbf{A}} {\mathbf{\Phi}}$ and least square fit inversion gives:

\begin{equation}
\mathbf{A}=\mathbf{g}\mathbf{\Phi}^T({\mathbf{\Phi}}{\mathbf{\Phi}}^T)^{-1},\end{equation}

where a singular value decomposition determines the pseudo-inverse of ${\mathbf{\Phi}}{\mathbf{\Phi}}^T$. The matrix $\mathbf{A}$ then gives the best-fit coefficients for the Hermite polynomial composition.

We furthermore adopt errors on $\mathbf{g}$ associated with Poisson counting, such that the uncertainty at each velocity coordinate $k$ is 
$\sigma_{g_{k}}= \sqrt {g_{k}}$ \citep{Wuest2007}. Poisson noise between energy bins is uncorrelated such that we construct the diagonal-weight matrix as $\mathbf{W}$ with entries  $\sigma^2$ can be included in weighted linear least square fitting as $\mathbf{gW}={\mathbf{A}} {\mathbf{\Phi}}\mathbf{W}$ such that
\begin{equation}\mathbf{A=gW\Phi}^T\mathbf{ (\Phi W} \mathbf{\Phi}^T)^{-1}\label{eq:wlsq}\end{equation}

The coefficients $\mathbf{A}$ are then used to construct $\bar{g}^H(v_\perp, v_\parallel)$ using Equation \ref{eq:hmte}.

Using $\bar{g}^H(v_\perp, v_\parallel)$, which is an analytic, differentiable funcion, it is possible to separately estimate $Q_{\rm ICW}$ without the assumption of a drifting bi-Maxwellian structure to the velocity distribution.

\subsection{Radial Basis Function Interpolation} We further include another approximation to the velocity distributions through interpolation via radial basis functions (RBFs) \citep{broomhead1988,Bowen2023} to model the distribution function via a summed set of interpolating functions given by
\begin{align} {g_p}^{RBF}(v_\perp, v_\parallel)= {\sum_{i}^{N_{RBF}-1}}R_i\psi_i(\zeta_i),\end{align} with $\zeta_i=\mathbf{v}- \mathbf{v}^c_i$, where $\mathbf{v}^c_i$ is the center of each interpolating RBF, denoted by $\psi_i$. The RBF method requires choice of a basis function, $\psi$, which we choose to be an isotropic 2D bi-Maxwellian
\begin{align} \psi_i(v_\perp, v_\parallel)= \frac{1}{\sqrt{\pi^{3}}w_{RBFi}^3}\text{exp}\left[-\frac{(v_\perp-v^c_{\perp i})^2}{w_{RBFi}^2} -\frac{(v_\parallel-v^c_{\parallel i})^2}{w_{RBFi}^2}\right].\end{align} 

The number of interpolating functions, $N_{RBF}$, must be specified along with the thermal speed $w_{{RBFi}}$ and central location, $\mathbf{v}^c_i$, of each of the $\psi_i$ implemented in the interpolation. In principle, $w_{RBFi}$ does not need to be isotropic and can vary for each $\psi_i$; however, we determine that setting a Maxwellian-RBF at each measured SPAN energy bin with nonzero phase space density gives suitable results. Furthermore, we uniformly set $w_{RBF}=40$ km/s, which we find is suitable for interpolating the proton distribution.

The interpolated velocity distribution via radial basis functions is then given by
\begin{align} \bar{g}^{RBF}(v_\perp, v_\parallel)= {\sum_{i}^{N_{RBF}-1}}R_i\psi_i(\zeta),\end{align} with $\zeta=\vec{v}- \vec{v}_c$. Determination of the weights $R_i$ is again performed through SVD estimation for the pseudoinverse giving a least square fit of $\bar{g}^{RBF}$ to the observed velocity distribution. We again compute the RBF on each 128s average velocity distribution and include weighted errors in the least square fit corresponding to Poisson noise analogous to Eq \ref{eq:wlsq}.

\subsection{Comparing Model Distributions}
To verify the heating of the plasma via ICW resonant interactions, we recompute Equation \ref{eq:H} using the Hermite polynomial and radial basis function approximations to the proton distribution, $\bar{g}^H(v_\perp, v_\parallel)$ and $\bar{g}^{RBF}(v_\perp, v_\parallel)$, we refer to the heating rates computed from Equation \ref{eq:H} via the drifting biMaxwellian, Hermite polynomials, and RBF interpolation as $Q_{BM}$, $Q_{HMTE}$, and $Q_{RBF}$ respectively, we also compute the average heating rate of the three terms $Q_{AVG}$. Note that the ICW spectrum $I(k_\parallel)$ in unaffected by the choice of model as $d_i=V_A/\omega_c$ is a function only of the magnetic field strength and total plasma density, which are independent of the kinetic phase-space distribution.

\begin{figure}
    \centering
    \includegraphics[width=.9\textwidth]{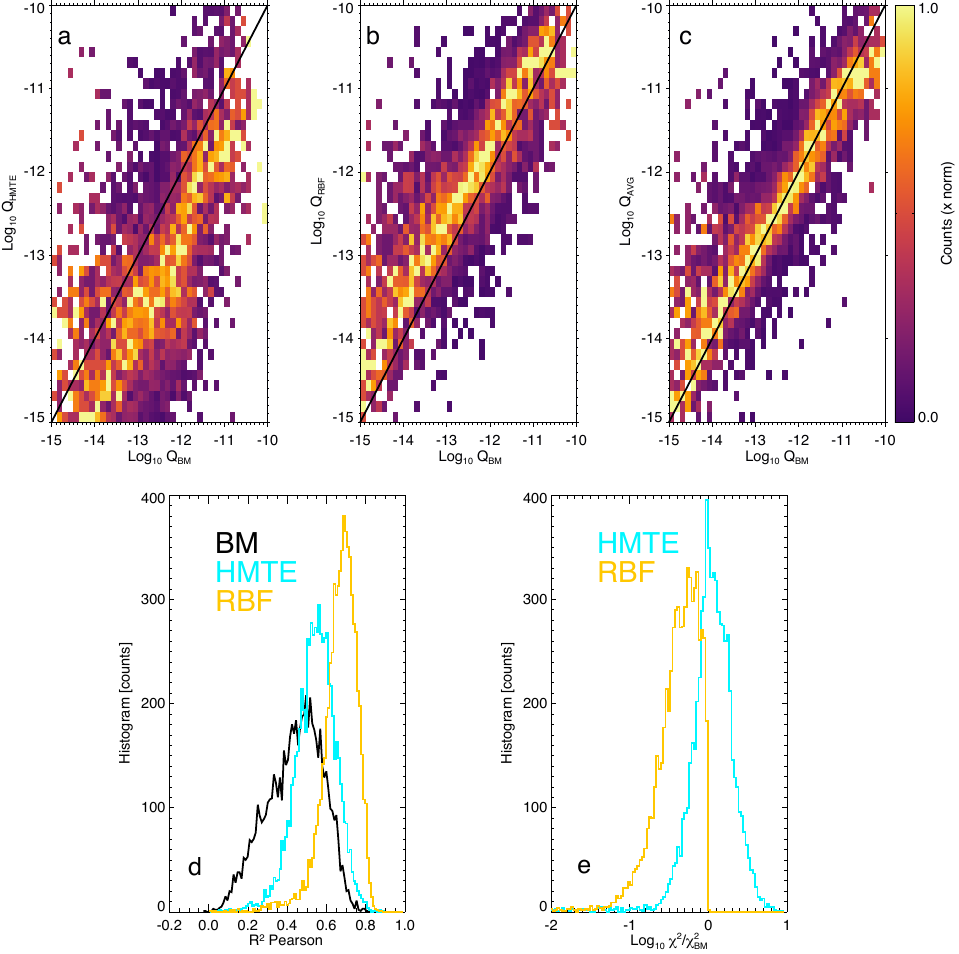}
    \caption{a) 2D histogram of $Q_{HMTE}$ against $Q_{BM}$ for each of the 6737 studied intervals; data are column normalized and the black line shows $Q_{HMTE}=Q_{BM}$. b) Same as panel a, but with $Q_{RBF}$ against $Q_{BM}$. c) Same as panel a, but with the average of all three heating rates $Q_{AVG}$ against $Q_{BM}$. d) Histogram showing distributions of Pearson $R^2$ correlation between the 6737 modeled distributions and observed SPANi distributions; drifting biMaxwellian (BM) in black; Hermite (HMTE) in teal; radial basis function (RBF) in orange. e) Histogram showing distributions of $\chi_{RBF}^2/\chi_{BM}^2$ (orange) and $\chi_{HMTE}^2/\chi_{BM}^2$ (teal) on a logarithmic scale.}
    \label{fig:Append}
\end{figure}
The top row of Figure \ref{fig:Append}, panels  (a-c), respectively show $Q_{HMTE}$, $Q_{RBF}$, and $Q_{AVG}$ against $Q_{BM}$. While all computed heating rates have the same qualitative behavior, i.e. positive valued indicating the absorption of ICWs and plasma heating via cyclotron resonance, $Q_{HMTE}$ tends to systematically underestimate the heating rates predicted by $Q_{BM}$, whereas $Q_{RBF}$ tends to overestimate $Q_{BM}$. The average of all heating rates $Q_{AVG}$ closely follows $Q_{BM}$, thus $Q_{BM}$ was chosen as the value reported in the body of the manuscript.

The bottom panels shows statistics on the quality of the various models. To compare the quality of the relative models we compute two parameters. Figure \ref{fig:Append}(d) shows the distribution of measured Pearson correlation coefficients between the logarithm of the observed distribution $\textrm{Log}_{10} g_p(v_\perp, v_\parallel)$ and the three models $\textrm{Log}_{10} \bar{g}^{BM}(v_\perp, v_\parallel)$, $\textrm{Log}_{10}\bar{g}^H(v_\perp, v_\parallel)$, and $\textrm{Log}_{10}\bar{g}^{RBF}(v_\perp, v_\parallel)$. High levels of correlation indicate that the model correctly matches the observed distribution. The use of the logarithm of the distribution in computing the Pearson $R^2$ correlation coefficient allows for weighting of the entirety of phase space and includes the regions where ICW resonant interactions are most important; without the logarithmic weighting, the correlation coefficient is mostly dominated by the models performance at the peak of the distribution. We find $R^2$ is largest for the RBF interpolation and smallest for the parametric drifting biMaxwellian fit. Interestingly, the level of heating computed between these models remains relatively constant even though the $R^2$ values vary, this suggests that variations between the three models are mostly occurring in regions that are not resonant with the outward going ICWs. We speculate these differences are likely due to the models ability to approximate the ion-beam population.

We also compute the $\chi^2$ of each of the models. Given the nature of the nonlinear fitting there is lack of clear number of degrees of freedom. Furthermore errors in the model are not likely to obey Gaussian statistics. For these reasons hypothesis testing via the $\chi^2$ statistic is not possible. To determine the quality of the fits we compare the ratio of $\chi^2$ to the value compared for the biMaxwellian, $\chi_{BM}^2$. Figure \ref{fig:Append}(e) shows the distribution of the logarithmic ratio of $\chi_{RBF}^2/\chi_{BM}^2$ and $\chi_{HMTE}^2/\chi_{BM}^2$. The RBF model tends to perform much netter than the drifting biMaxwellian with a uniform $\chi_{RBF}^2/\chi_{BM}^2<1$. In terms of $\chi^2$,  the Hermite polynomial approximation tends to have a similar performance to the drifting biMaxwellian fit in this stream, though the Pearson correlation is significantly better than the biMaxwellian fit. Again, the ability of each of models to reproduce similar heating rates, but with significantly different $\chi^2$ suggests that the main difference in the models may be in reproducing the ion-beam. Our future work will investigate further the role of the ion-beam in these processes.

In any case, the use of these various parametric and non-parametric  models for the observed proton distribution function all recover heating rates that suggest that ICW are absorbed into the core of the proton distribution, resulting in plasma heating via ICW resonance. These results suggest that the model used may not be inherently important in understanding ICW heating of the proton core.

\end{document}